\documentstyle[prd,aps,eqsecnum,twocolumn,floats]{revtex}

\def\beq{\begin{equation}}
\def\eeq{\end{equation}}
\def\beqa{\begin{eqnarray}}
\def\eeqa{\end{eqnarray}}
\def\bfig{\begin{figure}}
\def\efig{\end{figure}}

\input{psfig}
\begin{document}
\draft
\fnsymbol{footnote}

\wideabs{

\title{Gravitational waves from hot young rapidly rotating neutron stars}

\author{Benjamin J. Owen${}^1$, Lee Lindblom${}^1$, 
        Curt Cutler${}^2$, Bernard F. Schutz${}^2$, Alberto Vecchio${}^2$,
        and Nils Andersson${}^3$}

\address{${}^1$Theoretical Astrophysics 130-33, California Institute
         of Technology, Pasadena, California 91125} 

\address{${}^2$Max Planck Institute for Gravitational Physics,
        Schlaatsweg 1, D-14473 Potsdam, Germany}
         
\address{${}^3$Institute for Astronomy and Astrophysics,
         University of T\"ubingen, D-72076 T\"ubingen, Germany}

\date{\today}
\maketitle

\begin{abstract}

Gravitational radiation drives an instability in the {\it r-}modes of
young rapidly rotating neutron stars.  This instability is expected to
carry away most of the angular momentum of the star by gravitational
radiation emission, leaving a star rotating at about 100~Hz.  In this
paper we model in a simple way the development of the instability and
evolution of the neutron star during the year-long spindown phase.
This allows us to predict the general features of the resulting
gravitational waveform.  We show that a neutron star formed in the
Virgo cluster could be detected by the LIGO and VIRGO gravitational
wave detectors when they reach their ``enhanced'' level of
sensitivity, with an amplitude signal-to-noise ratio that could be as
large as about 8 if near-optimal data analysis techniques are
developed.  We also analyze the stochastic background of gravitational
waves produced by the $r$-mode radiation from neutron-star formation
throughout the universe.  Assuming a substantial fraction of neutron
stars are born with spin frequencies near their maximum values, this
stochastic background is shown to have an energy density of about
$10^{-9}$ of the cosmological closure density, in the range 20~Hz to
1~kHz.  This radiation should be detectable by ``advanced'' LIGO as well.
 
\pacs{PACS Numbers: 04.30.Db, 04.40.Dg, 97.60.Jd}
\end{abstract}
}

\narrowtext
\section{Introduction}
\label{sectionI}

Recently Andersson \cite{andersson} discovered that gravitational
radiation tends to destabilize the {\it r-}modes of rotating stars.
Friedman and Morsink \cite{friedman-morsink} then showed that this
instability is generic, in the sense that gravitational radiation
tends to make {\it all r-}modes in {\it all} rotating stars unstable.
Lindblom, Owen, and Morsink \cite{lindblom-owen-morsink} have recently
evaluated the timescales associated with the growth of this
instability.  Gravitational radiation couples to these modes through
the current multipoles rather than the more typical mass multipole
moments.  This coupling is stronger than anyone anticipated for these
modes, and is so strong in fact that the viscous forces present in hot
young neutron stars are not sufficient to suppress the gravitational
radiation driven instability.  Gravitational radiation is expected
therefore to carry away most of the angular momentum of hot young
neutron stars.  These results have now been verified by
Andersson, Kokkotas, and Schutz~\cite{andersson-et-al}.

In this paper we study the gravitational waveforms that are produced
as the {\it r-}mode instability grows and radiates away the bulk of
the angular momentum of a hot young rapidly rotating neutron star.
The properties of the {\it r-}modes and the instability associated
with them are reviewed in Sec.~\ref{sectionII}.  The equations that
describe (approximately) the evolution of the {\it r-}modes as they
grow and spin down a rapidly rotating neutron star are derived in
Sec.~\ref{sectionIII}.  These equations are solved numerically and the
results are also presented in Sec.~\ref{sectionIII}.  The
gravitational waveforms associated with the {\it r}-mode instability
are evaluated in Sec.~\ref{sectionIV}.  General analytical and
detailed numerical expressions for these waveforms are presented.  In
Sec.~\ref{sectionV} we evaluate the detectability of this type of
gravitational wave signal by the laser interferometer gravitational
wave detectors such as LIGO~\cite{science}, VIRGO~\cite{virgo}, and
GEO~\cite{geo600}.  We consider the detectability of signals produced
by single nearby sources, and also the detectability of a stochastic
background of sources from throughout the universe.  Finally, in
Sec.~\ref{secVI} we discuss the prospects for gravitational-wave
astronomy opened up by the $r$-modes.

\section{The {\it r}-Mode Instability}
\label{sectionII}

The {\it r-}modes of rotating Newtonian stars are generally defined to
be solutions of the perturbed fluid equations having (Eulerian)
velocity perturbations of the form

\beq \delta \vec{v} = R \Omega f(r/R) \vec{Y}^{B}_{l\,m}
e^{i\omega t},\label{2.1} \eeq

\noindent where $R$ and $\Omega$ are the radius and angular velocity
of the unperturbed star, $f(r/R)$ is an arbitrary dimensionless
function, and $\vec{Y}^{B}_{l\,m}$ is the magnetic type vector
spherical harmonic defined by

\beq
\vec{Y}^{B}_{l\,m}= [l(l+1)]^{-1/2}r\vec{\nabla}
\times(r \vec{\nabla}Y_{l\,m}).
\label{2.2}
\eeq
 
\noindent For barotropic stellar models, of primary concern to us
here, the Euler equation determines the form of these modes:
The radial dependence $f(r/R)$ is determined to be $f(r/R)=\alpha
(r/R)^l$, where $\alpha$ is an arbitrary constant~\cite{provost}.
These modes exist with velocity perturbations as given by
Eq.~(\ref{2.1}) if and only if $l=m$~\cite{provost}.  Also, the
frequencies of these modes are given by~\cite{papaloizou-pringle}

\beq
\omega = - {(l-1)(l+2)\over l+1}\Omega.\label{2.3}
\eeq

\noindent These modes represent large scale oscillating currents that
move (approximately) along the equipotential surfaces of the rotating
star.  The restoring force for these oscillations is the Coriolis
force; hence the frequencies of these modes are low compared to the
usual $f$ and $p-$modes in slowly rotating stars.  These expressions
for $\delta \vec{v}$ and $\omega$ are the lowest order terms in an
expansion in terms of the angular velocity $\Omega$.  The exact
expressions contain additional terms of order $\Omega^3$.  There may
exist other modes of rotating barotropic stellar models with
properties similar to these classical {\it r-}modes; however, our
discussion here is limited to the properties of these classical {\it
r-}modes.

The density perturbation associated with the {\it r-}modes can be
deduced by evaluating the inner product of $\vec{v}$ (the unperturbed
fluid velocity) with the perturbed Euler equation, and the equation
for the perturbed gravitational potential \cite{ipser-lind}:
 
\beqa
\delta \rho &&=
\alpha R^2\Omega^2 \rho {d\rho\over dp} 
\left[{2 l\over 2l+1}\sqrt{l\over l+1}
\left({r\over R}\right)^{l+1}+\delta\Psi(r)\right]\nonumber\\
&&\qquad\qquad\qquad\times Y_{l+1\,l}
\,e^{i\omega t}.\label{2.4}
\eeqa

\noindent The quantity $\delta\Psi$ is proportional to the perturbed
gravitational potential $\delta\Phi$, and is the solution to the
ordinary differential equation

\beqa
{d^2\delta \Psi(r)\over dr^2} &&+ {2\over r} {d\delta \Psi(r)\over dr}
+\left[4\pi G\rho {d\rho\over dp}- {(l+1)(l+2)\over r^2}\right]\delta
\Psi(r)
\nonumber\\&&\qquad
= -  {8\pi G l\over 2l+1} \sqrt{l\over l+1}
\rho {d\rho\over dp}\left({r\over R}\right)^{l+1},\label{2.5}
\eeqa
 
\noindent which satisfies appropriate asymptotic boundary conditions.
We note that $\delta \rho$ is proportional to $\Omega^2$ and hence is
small (i.e., higher order in $\Omega$) compared to $\delta\vec{v}$ in
slowly rotating stars.  We also note that $\delta\rho$ is proportional
to $Y_{l+1\,l}$---having spherical harmonic index one order in $l$
higher than that of the velocity perturbation.  Equation~(\ref{2.4})
is the complete expression for the density perturbation to order
$\Omega^2$.  The exact expression for $\delta\rho$ includes additional
terms of order $\Omega^4$.

The {\it r-}modes evolve with time dependence $e^{i\omega t - t/\tau}$
as a consequence of ordinary hydrodynamics and the influence of the
various dissipative processes.  The real part of the frequency of these
modes, $\omega$, is given in Eq.~(\ref{2.3}), while the imaginary part
$1/\tau$ is determined by the effects of gravitational radiation,
viscosity, etc.  The simplest way to evaluate $1/\tau$ is to compute the
time derivative of the energy $\tilde{E}$ of the mode (as measured in
the rotating frame). $\tilde{E}$ can be expressed as
a real quadratic functional of the fluid perturbations:

\beq
\tilde{E}={1\over 2}\int\left[\rho\delta\vec{v}\cdot\delta\vec{v}^*
+\left({\delta p\over \rho}-\delta\Phi\right)\delta\rho^*\right]d^3x.
\label{2.6}
\eeq

\noindent Thus the time derivative of $\tilde{E}$ is related to the
imaginary part of the frequency $1/\tau$ by

\beq
{d\tilde{E}\over dt}= -{2\tilde{E}\over \tau}.
\label{2.7}
\eeq

\noindent Since the specific expressions for the time derivative of
$\tilde{E}$ due to the influences of gravitational radiation
\cite{thorne} and viscosity \cite{ipser-lindblom91} are well known,
Eq.~(\ref{2.7}) may be used to evaluate the imaginary part of the
frequency.

It is convenient to decompose $1/\tau$:

\beq
{1\over \tau(\Omega)} = 
{1\over \tau_{\scriptscriptstyle GR}(\Omega)}
+ {1\over \tau_{\scriptscriptstyle S}(\Omega)}
+ {1\over \tau_{\scriptscriptstyle B}(\Omega)},
\label{2.8}
\eeq

\noindent where ${1/ \tau_{\scriptscriptstyle GR}}$, ${1/
\tau_{\scriptscriptstyle S}}$, and ${1/ \tau_{\scriptscriptstyle B}}$
are the contributions due to gravitational radiation emission, shear
viscosity and bulk viscosity respectively.  Expressions for these
individual contributions for the $r-$modes are given by
\cite{lindblom-owen-morsink}:

\beqa
{1\over \tau_{\scriptscriptstyle GR}} 
=-{32\pi G\Omega^{2l+2}\over c^{2l+3}}&&{ (l-1)^{2l}\over [(2l+1)!!]^2}
\left({l+2\over l+1}\right)^{2l+2}\nonumber\\
&&\times\int_0^R\rho\,r^{2l+2} dr,\label{2.9}
\eeqa

\beq
{1\over \tau_{\scriptscriptstyle S}} 
= (l-1)(2l+1) \int_0^R \eta r^{2l} dr
\left( \int_0^R \rho\, r^{2l+2} dr\right)^{-1}.
\label{2.10}
\eeq

\noindent and

\beq
{1\over \tau_{\scriptscriptstyle B}}
\approx {4 R^{2l-2}\over (l+1)^2}\int\zeta
\left|{\delta\rho\over\rho}\right|^2d^3x\left(\int_0^R
\rho\,r^{2l+2}dr\right)^{-1},\label{2.11}
\eeq

\noindent where $\delta\rho$ is given in Eq.~(\ref{2.4}).  We note
that the expression for $1/\tau_{\scriptscriptstyle B}$ in
Eq.~(\ref{2.11}) is only approximate.  The exact expression should
contain the Lagrangian density perturbation $\Delta\rho$ in place of
the Eulerian perturbation $\delta\rho$.  The bulk viscosity (see
Eq.~\ref{2.13}) is a very strong function of the temperature, being
proportional to $T^6$.  Thus, the result of any error that might occur
in our approximation for $1/\tau_{\scriptscriptstyle B}$ is simply to
shift slightly the temperature needed to achieve a given viscosity
timescale.  Numerical estimates show that changing this quantity by
even a factor of one hundred (as suggested by
Ref.~\cite{andersson-et-al}), does not substantially affect the
important physical quantities computed here (i.e.~the spindown rate or
the final angular velocity of the star).

We have evaluated these expressions for the imaginary parts of the
frequency for a ``typical'' neutron star model with a polytropic
equation of state: $p=k\rho^2$, with $k$ chosen so that a
1.4$M_\odot$ model has the radius $12.53$km.  We use the usual
expressions for the viscosity of hot neutron star matter \cite{visrefs}:

\beq
\eta=347\rho^{9/4}T^{-2},\label{2.12}
\eeq

\beq
\zeta=6.0\times 10^{-59}
\left({l+1\over 2\Omega}\right)^2\rho^2T^6,\label{2.13}
\eeq

\noindent where all quantities are expressed in cgs units.  We have
evaluated the expressions, Eqs.~(\ref{2.9})--(\ref{2.11}), for the
dissipative timescales with fiducial values of the angular velocity
$\Omega=\sqrt{\pi G \bar{\rho}}$ and temperature $T=10^9$K.  These
fiducial timescales ${\tilde{\tau}_{\scriptscriptstyle GR}}$,
${\tilde{\tau}_{\scriptscriptstyle V}}$, and
${\tilde{\tau}_{\scriptscriptstyle B}}$ are given in
Table~\ref{table1} for the {\it r-}modes with $2\leq l \leq 6$.  It
will be useful in the following to define a timescale associated with
the viscous dissipation $1/\tau_{\scriptscriptstyle
V}=1/\tau_{\scriptscriptstyle S} +1/\tau_{\scriptscriptstyle B}$.  The
viscous timescale $\tau_{\scriptscriptstyle V}$ and the gravitational
timescale $\tau_{\scriptscriptstyle GR}$ can be expressed then in terms
of the fiducial timescales in a way that makes their temperature and
angular velocity dependences explicit:

\beq
{1\over \tau_{\scriptscriptstyle V}}=
{1\over  \tilde{\tau}_{\scriptscriptstyle S}}
\left({10^9 {\rm K}\over T}\right)^2
+ {1\over  \tilde{\tau}_{\scriptscriptstyle B}}
\left({T\over 10^9 {\rm K}}\right)^6
\left({\Omega^2\over \pi G \bar{\rho}}\right),
\label{2.14}
\eeq

\beq
{1\over\tau_{\scriptscriptstyle GR}} = 
{1\over  \tilde{\tau}_{\scriptscriptstyle GR}}
\left({\Omega^2\over \pi G \bar{\rho}}\right)^{l+1}.
\label{2.15}
\eeq

\begin{table}[b]
\caption{Gravitational radiation and viscous timescales (in seconds)
are presented for $T=10^9\,$K and $\Omega=\sqrt{\pi G \bar{\rho}}$.}
\begin{tabular}{cccc}
$\qquad l\qquad $&$\qquad \tilde{\tau}_{\scriptscriptstyle GR}\qquad$
&$\qquad\tilde{\tau}_{\scriptscriptstyle S}\qquad$
&$\qquad\tilde{\tau}_{\scriptscriptstyle B}\qquad$\\
\hline
2&$-3.26\times 10^0$ & $2.52\times 10^8$ & $6.99\times 10^8$\\
3&$-3.11\times 10^1$ & $1.44\times 10^8$ & $5.13\times 10^8$\\
4&$-2.85\times 10^2$ & $1.07\times 10^8$ & $4.01\times 10^8$\\
5&$-2.37\times 10^3$ & $8.79\times 10^7$ & $3.26\times 10^8$\\
6&$-1.82\times 10^4$ & $7.58\times 10^7$ & $2.74\times 10^8$
\end{tabular}
\label{table1}
\end{table}

\section{Evolution of the {\it r-}Modes}
\label{sectionIII}

To determine the gravitational waveform that will result from the
instability in the {\it r-}modes, we must estimate how the neutron
star evolves as the instability grows and radiates the angular momentum
of the star away to infinity.  Initially the mode will be a small
perturbation that is described adequately by the linear analysis that
we have described above.  However, as the mode grows, non-linear
hydrodynamic effects become important and eventually dominate the
dynamics.  At the present time we do not have available the tools to
follow exactly this non-linear phase of the evolution.  Instead, we
propose a simple approximation that includes (we believe) the basic
features of the exact evolution.

We treat the star as a simple system having only two degrees of
freedom: the uniformly rotating equilibrium state parameterized by its
angular velocity $\Omega$, and the {\it r-}mode parameterized by its
amplitude $\alpha$.  The total angular momentum $J$ of this simple
model of the star is given by,

\beq
J = I\Omega + J_c,\label{3.1}
\eeq

\noindent where $I$ is the moment of inertia of the equilibrium state
of the star, and $J_c$ is the canonical angular momentum of the {\it
r-}mode.

In this simple model of the star the angular momentum $J$ is a
function of the two parameters that characterize the state of the
system: $J=J(\Omega,\alpha)$.  We can determine this functional
relationship approximately as follows.  The canonical angular momentum
of an {\it r-}mode can be expressed in terms of the velocity
perturbation $\delta\vec{v}$ by \cite{friedman-schutz},

\beq
J_c = -{l\over2 (\omega+l\Omega)}\int \rho\,
\delta\vec{v}\cdot\delta\vec{v}^{\,*} d^3x.
\label{3.2}
\eeq

\noindent For the $l=2$ {\it r-}mode of primary
interest to us here this expression reduces (at lowest
order in $\Omega$) to

\beq
J_c=-{\scriptstyle \frac{3}{2}}\Omega\alpha^2\tilde{J}MR^2,\label{3.3}
\eeq

\noindent where $\tilde{J}$ is defined by

\beq
\tilde{J}={1\over MR^4}\int_0^R\rho\, r^{6}dr.\label{3.4}
\eeq

\noindent For the polytropic models studied in detail here the 
dimensionless constant $\tilde{J}=1.635\times 10^{-2}$.  The
moment of inertia $I$ can also be conveniently expressed as

\beq
I = \tilde{I}MR^2,\label{3.5}
\eeq

\noindent where $\tilde{I}$ is given by

\beq
\tilde{I} = {8\pi\over 3 MR^2}\int_0^R\rho\,r^4dr.\label{3.6}
\eeq

\noindent For the polytropic models considered here $\tilde{I}= 0.261$.
Thus, our simple model of the angular momentum of the perturbed star is

\beq
J(\Omega,\alpha)= (\tilde{I}
-{\scriptstyle \frac{3}{2}}\tilde{J}\alpha^2)\Omega MR^2.
\label{3.7}
\eeq

The perturbed star loses angular momentum primarily through the
emission of gravitational radiation.  Thus, we compute the evolution
of $J(\Omega,\alpha)$ by using the standard multipole expression for
angular momentum loss.  The $l=2$ {\it r-}mode is the primary source
of gravitational radiation in our simple model of this system, and
this mode loses angular momentum primarily through the $l=m=2$ current
multipole.  Thus the angular momentum of the star evolves as

\beq
{dJ\over dt} = - {c^3\over 16\pi G} \left({4\Omega\over 3}\right)^5
(S_{22})^2.\label{3.8}
\eeq

\noindent The $l=m=2$ current multipole moment $S_{22}$ for this
system is given by

\beq
S_{22} = \sqrt{2}{32\pi\over 15}{GM\over c^5}\alpha \Omega R^3\tilde{J}
.\label{3.9}
\eeq

\noindent Combining Eq.~(\ref{3.8}) for the angular momentum evolution
of the star with Eqs.~(\ref{2.9}), and (\ref{3.7}), we
obtain one equation for the evolution of the parameters $\Omega$ and
$\alpha$ that determine the state of the star:

\beq
(\tilde{I}-{\scriptstyle \frac{3}{2}}\alpha^2\tilde{J}){d\Omega\over dt}
-3\alpha\Omega\tilde{J}{d\alpha\over dt}={3\alpha^2\Omega\tilde{J}\over
{\tau}_{\scriptscriptstyle GR}}.
\label{3.10}
\eeq

During the early part of the evolution of the star, the perturbation
analysis of the {\it r-}modes described earlier applies.  In addition
to radiating angular momentum from the star via gravitational
radiation, the mode will also lose energy via gravitational radiation
and neutrino emission (from the bulk viscosity) and also deposit
energy into the thermal state of the star due to shear viscosity.  It
is most convenient to obtain the equation for the energy balance
during this part of the evolution in terms of the energy $\tilde{E}$
of the mode as defined in Eq.~(\ref{2.6}).  For the $l=2$ {\it r-}mode 
$\tilde{E}$ is given by

\beq
\tilde{E}=
{\scriptstyle \frac{1}{2}}\alpha^2\Omega^2 MR^2\tilde{J}.
\label{3.11}
\eeq

\noindent The time derivative of $\tilde{E}$ is precisely the quantity
that was used to determine the imaginary part of the frequency
of the mode in Eq.~(\ref{2.7}):

\beq
{d\tilde{E}\over dt}=-2\tilde{E}\left(
{1\over \tau_{\scriptscriptstyle GR}}+
{1\over \tau_{\scriptscriptstyle V}}\right)
.\label{3.12}
\eeq

\noindent Equation~(\ref{3.12}) together with (\ref{3.11})
therefore provides a second equation
for determining the evolution of the parameters $\Omega$ and $\alpha$
that specify the state of the star:

\beq
\Omega{d\alpha\over dt} + \alpha {d\Omega\over dt}=-\alpha\Omega
\left(
{1\over \tau_{\scriptscriptstyle GR}}+
{1\over \tau_{\scriptscriptstyle V}}\right)
.\label{3.13}
\eeq

\noindent Equations~(\ref{3.10}) and (\ref{3.13}) can be combined then to
determine the evolution of $\Omega$ and $\alpha$ during the portion of
the evolution in which the perturbation remains small:

\beq
{d\Omega\over dt}=-{2\Omega\over \tau_{\scriptscriptstyle V}}
{\alpha^2 Q\over 1+\alpha^2Q}
,\label{3.14}
\eeq

\beq
{d\alpha\over dt}=-{\alpha\over\tau_{\scriptscriptstyle GR}}
-{\alpha\over\tau_{\scriptscriptstyle V}}{1-\alpha^2Q\over 1+\alpha^2Q}
.\label{3.15}
\eeq

\noindent The equation of state dependent parameter $Q$ that appears in
Eqs.~(\ref{3.14}) and (\ref{3.15}) is defined by $Q=3\tilde{J}/2\tilde{I}$.
For the polytropic model considered in detail here $Q=9.40\times10^{-2}$.
We note that during the initial linear evolution phase the
angular velocity of the star $\Omega$ is nearly constant, evolving
according to Eq.~(\ref{3.14}) on the viscous dissipation timescale. 
During this phase the amplitude of the mode $\alpha$ grows
exponentially on a timescale that is comparable to the
gravitational radiation timescale. 

After a short time (about $500\,$s in our numerical solutions) the
amplitude becomes so large that non-linear effects can no longer be
ignored.  We have not yet developed the tools needed to follow the
evolution exactly during this non-linear phase.  However, we do have
some intuition about the non-linear hydrodynamical evolution of
gravitationally driven instabilities in rotating stars.  This
intuition comes from the studies of the effects of gravitational
radiation reaction on the evolution of the ellipsoidal models
\cite{detweiler-lindblom,lai-shapiro}.  In that case the unstable mode
grows exponentially until its amplitude is of order unity.  At that
point a kind of non-linear saturation occurs, and the growth of the
mode stops.  The excess angular momentum of the star is radiated away
and the star evolves toward a new lower angular momentum equilibrium
state.  We expect a similar situation to pertain in the evolution of
the {\it r-}modes.  Thus, we expect non-linear effects will saturate
and halt the further growth of the mode when the amplitude of the mode
becomes of order unity.  Thus, when the amplitude $\alpha$ grows to
the value

\beq
\alpha^2=\kappa,
\label{3.16}
\eeq

\noindent (where $\kappa$ is a constant of order unity) we stop
evolving the star using Eqs.~(\ref{3.14}) and (\ref{3.15}).  Instead we
set $d\alpha/dt=0$ during the saturated non-linear phase of the
evolution, while continuing to evolve the angular velocity $\Omega$ by
Eq.~(\ref{3.10}) as angular momentum is radiated away to infinity by
gravitational radiation.  During this phase, then the angular velocity
evolves by

\beq
{d\Omega\over dt}
= {2\Omega\over \tau_{\scriptscriptstyle GR}}
{\kappa Q\over 1-\kappa Q}.
\label{3.17}
\eeq

The {\it r-}mode will evolve during the saturated non-linear phase of
its evolution approximately according to Eqs.~(\ref{3.16}) and
(\ref{3.17}).  During this phase the star will lose most of its
angular momentum, and spin down to a state having an angular velocity
that is much smaller than $\Omega_{\scriptstyle K}\approx
{\scriptstyle \frac {2}{3}}\sqrt{\pi G \bar{\rho}}$.  The star will
eventually (in about 1 year in our numerical solutions) evolve to a
point where the angular velocity and temperature become sufficiently
low that the {\it r-}mode is no longer unstable.  The end of the evolution
is characterized by a phase in which the viscous forces and
gravitational radiation damp out the energy remaining in the mode and
move the star slowly to its final equilibrium configuration.  During
this final phase, the mode is again of small amplitude and so the
linear approximation is adequate to describe the evolution.  We
monitor the quantity on the right side of Eq.~(\ref{3.15}) throughout
the non-linear evolution phase.  When it becomes negative we change
the evolution equations again, from Eqs.~(\ref{3.16}) and (\ref{3.17})
back to the linear equations Eqs.~(\ref{3.14}) and (\ref{3.15}).

In summary then, we model the evolution of the {\it r-}mode as having
three distinct phases: (i)~The hot young neutron star is born rapidly
rotating with a small initial excitation in the $l=2$ {\it r-}mode.
This mode initially grows exponentially according to Eqs.~(\ref{3.14})
and (\ref{3.15}). (ii)~The amplitude of the mode saturates due to
non-linear hydrodynamic effects at a value of order unity.  The bulk
of the angular momentum of the star is radiated away by gravitational
radiation during this phase according to Eqs.~(\ref{3.16}) and
(\ref{3.17}). (iii)~The final phase of the evolution begins when the right
side of Eq.~(\ref{3.15}) becomes negative so that the mode begins to be
damped out.  During the final phase the star evolves again according
to Eqs.~(\ref{3.14}) and (\ref{3.15}).

In order to complete our model for the evolution of the {\it r-}modes
we must specify how the temperature of the star evolves with time.  We
do this by adopting one of the standard descriptions of the cooling of
hot young neutron stars.  These stars are expected to cool primarily
due to the emission of neutrinos via a modified URCA process.  The
temperature during this phase falls quickly by a simple power law
cooling formula \cite{shapiro-teukolsky}:

\beq
{T(t)\over 10^9K}=\left[ {t\over \tau_c} + \left({10^9K\over
T_i}\right)^6
\right]^{-1/6},\label{3.18}
\eeq 

\noindent where $T_i$ is the initial temperature of the neutron star,
and $\tau_c$ is a parameter that characterizes the cooling rate.  For
the modified URCA process $\tau_c\approx 1$y.  A typical value for the
initial temperature is $T_i\approx 10^{11}$K.  Equation~(\ref{3.18})
can now be inserted into Eqs.~(\ref{3.14})--(\ref{3.17}) to provide
explicit differential equations for the time evolution of the angular
velocity of the star and the amplitude of the mode.  These equations
can be solved numerically in a straightforward manner.

Figs.~\ref{fig1} and \ref{fig2} illustrate the solutions to these
equations.  The dashed curves in Figs.~\ref{fig1} and~\ref{fig2} show
the critical angular velocity $\Omega_c$, defined by
$1/\tau(\Omega_c)=0$, above which the {\it r-}modes are unstable.
Figure~\ref{fig1} shows the evolution ($\Omega$ plotted versus $T$) of
the angular velocity of the star for $\kappa=1.0$ and a range of
values of the initial value of the parameter $\alpha$.  In these
simulations we have assumed that the initial angular velocity of the
star is $\Omega=\Omega_{\scriptstyle K}$.  This figure illustrates
that the final non-linear part of the evolution is remarkably
insensitive to the initial size of the perturbation.

\bfig
\centerline{\psfig{file=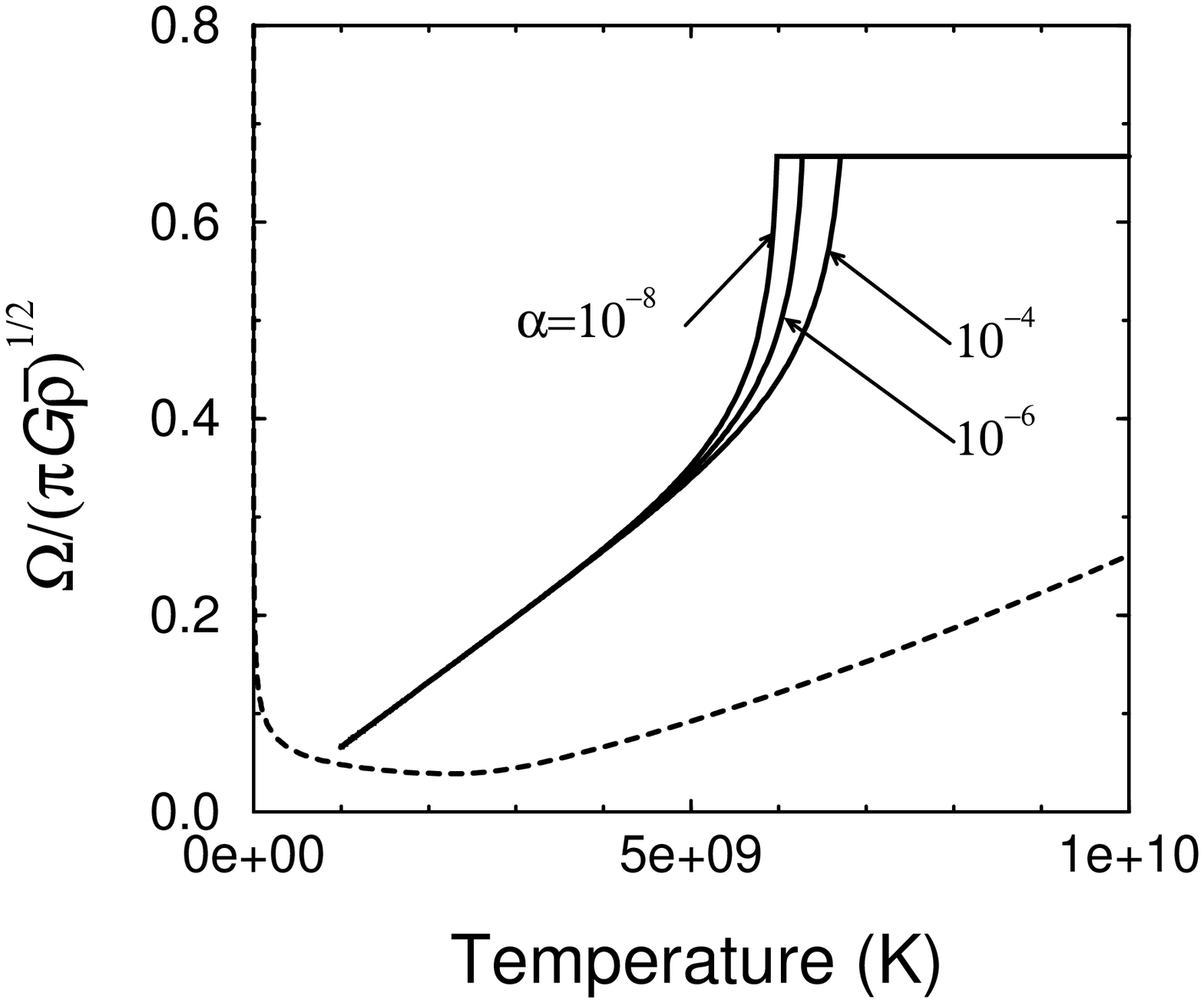,height=2.0in}}
\vskip 0.3cm
\caption{Angular velocity evolution as gravitational radiation
spins down a hot rapidly rotating neutron star.  $\alpha$ measures the
amplitude of the initial perturbation. The dashed curve shows the
critical angular velocity $\Omega_c$ above which the {\it r-}modes
are unstable.
\label{fig1}
}
\efig

Figure~\ref{fig2} illustrates the dependence of the evolution on the
parameter $\kappa$ by showing several evolutions with initial values
of $\alpha=10^{-6}$.  The parameter $\kappa$ measures the degree of
saturation that occurs in the non-linear spindown phase of the star.
In our numerical studies we examine the limited range
$0.25\leq\kappa\leq 2$. If $\kappa$ is taken to be too small, the mode
simply does not grow to the point that non-linear effects can stop its
growth.  Conversely if $\kappa$ is taken too large, then our simple
evolution equations based in part on the linear perturbation theory
become singular, e.g.\ Eq.~(\ref{3.17}).  The equations break down
because the (negative) canonical angular momentum of the mode equals
the (positive) angular momentum of the equilibrium configuration, and
therefore Eq.~(\ref{3.1}) yields the unphysical result that the star
has no net angular momentum.

We artificially stop all of our evolution curves when the temperature
of the star falls to $10^9\,$K.  Below this temperature we expect
superfluidity and perhaps other non-perfect fluid effects to make our
simple simulation highly inaccurate~\cite{lindblom-mendell}.
Figure~\ref{fig2} illustrates that the gravitational radiation
instability in the {\it r-}modes is nevertheless effective in
radiating away most of the angular momentum of the star before the
star cools to the point that superfluidity or other effects are
expected to become important.  Figure~\ref{fig2} shows that the amount
of angular momentum lost in this process is remarkably insensitive to
the value of $\kappa$.  Thus, the final upper limit on the angular
velocity of the star is (fortunately) fairly insensitive to our
assumption about the exact nature of the non-linear portion of the
star's evolution.

\bfig
\centerline{\psfig{file=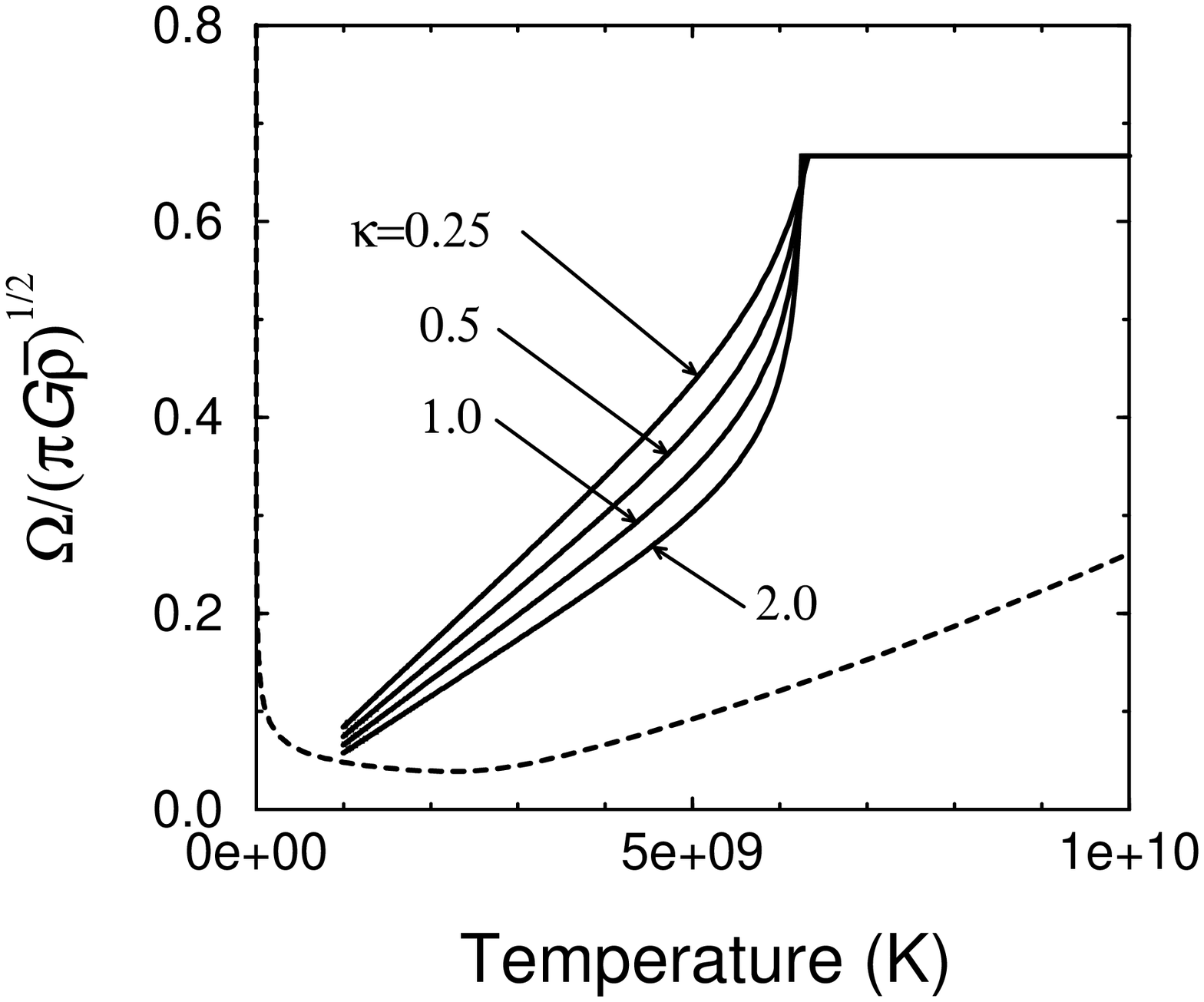,height=2.0in}}
\vskip 0.3cm
\caption{Evolution of the angular velocity of the star depends 
on the parameter $\kappa$ in the non-linear saturated phase, but the
final angular velocity of the star is insensitive to
this. The dashed curve is the same as in Fig. 1.\label{fig2}}
\efig

In this simple model of the evolution of the unstable star, we have
ignored the effect that viscous heating might have on the cooling rate
of the star.  If there were too much viscous heating, then the cooling
formula given in Eq.~(\ref{3.18}) would not be correct.  We have
evaluated the importance of this re-heating effect by comparing the
rate at which thermal energy is being radiated away from the star by
neutrinos according to Eq.~(\ref{3.18}) with the rate that viscous
dissipation deposits thermal energy into the star.  Neutrino cooling
removes energy from the thermal state of the star at the rate
\cite{shapiro-teukolsky}

\beq
{dU\over dt}= 7.4\times 10^{39}\left({T\over 10^9{\rm K}}\right)^8
{\rm ergs/s}.\label{3.19}
\eeq

\noindent Thermal energy is generated by shear viscosity as the star
evolves, but not by bulk viscosity.  Bulk viscosity radiates away its
excess energy directly by neutrinos without significantly interacting
with the thermal energy contained in the star.  Thus, energy is
transfered from the canonical energy of the {\it r-}mode to the
thermal energy of the star by the formula

\beq
{dE_c\over dt}={2\alpha^2\Omega^2 MR^2 \tilde{J}\over
{\tau}_{\scriptstyle S}}.
\label{3.20}
\eeq

Figure~\ref{fig3} compares the values of $dU/dt$ and $dE_c/dt$ for our
numerical evolution with $\kappa=1.0$ and initial $\alpha=10^{-6}$.
We find that viscosity does not deposit energy into the thermal state
of the star at a significant rate until the temperature of the system
falls to about $10^9\,$K.  At this point the star has already lost
most of its angular momentum to gravitational radiation, and other
dissipative effects (such as those associated with superfluidity)
which are not modeled here will start to play a significant
role~\cite{lindblom-mendell}.  Thus, we are justified in ignoring the
effects of viscous re-heating on the thermal evolution of the star
during the early part of its evolution modeled here.

\bfig
\centerline{\psfig{file=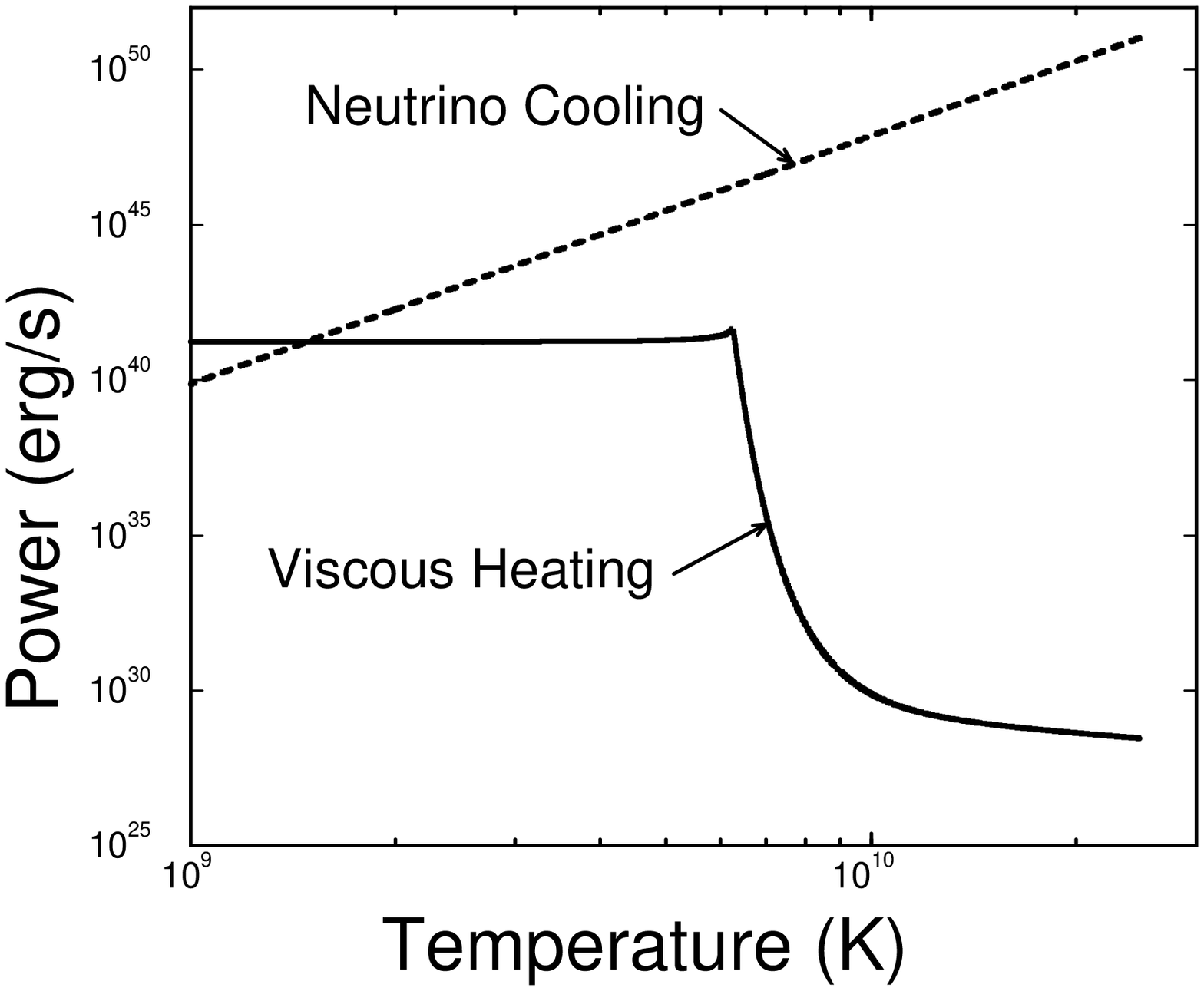,height=2.0in}}
\nobreak\vskip 0.3cm\nobreak
\caption{Cooling rate due to neutrino emission is compared to the
viscous heating rate in our numerical evolution of the unstable {\it
r-}modes.
\label{fig3}}
\efig

The modified URCA process that determines the thermal evolution used
in our evolutions is the standard mechanism by which neutron stars
are expected to cool down to about $10^9\,$K.
Other less standard mechanisms have also been proposed which could
significantly speed up the cooling \cite{shapiro-teukolsky}.  These
mechanisms include neutrino emission processes that require the
presence of exotic species (such as quarks or pions) as free particles
in the cores of these stars.  We have ignored these possibilities in
the evolutions described above.  If these particles do exist in the
cores of neutron stars, we expect that they will only be present in a
small volume of material at the centers of these stars.  The cores of
these stars may well cool rapidly, but the outer layers where the {\it
r-}mode is large will continue to cool at the rate given in
Eq.~(\ref{3.18}) until thermal conduction can move energy from the outer
layers back into the core.

To estimate what effect a somewhat more rapid cooling might have on
the evolution of the {\it r-}modes, we have artificially varied the
value of the parameter $\tau_c$ that determines the cooling rate in
Eq.~(\ref{3.18}).  Figure~\ref{fig4} shows the results of the evolution
of an {\it r-}mode with initial $\alpha=10^{-6}$ and $\kappa=1.0$ for
several values of $\tau_c$.  We see that while the details of the
evolution are somewhat effected, the total amount of angular momentum
radiated away by gravitational radiation, and the final angular
velocity of the star are fairly insensitive to the rate at which the
star cools to $10^9\,$K.

\bfig
\centerline{\psfig{file=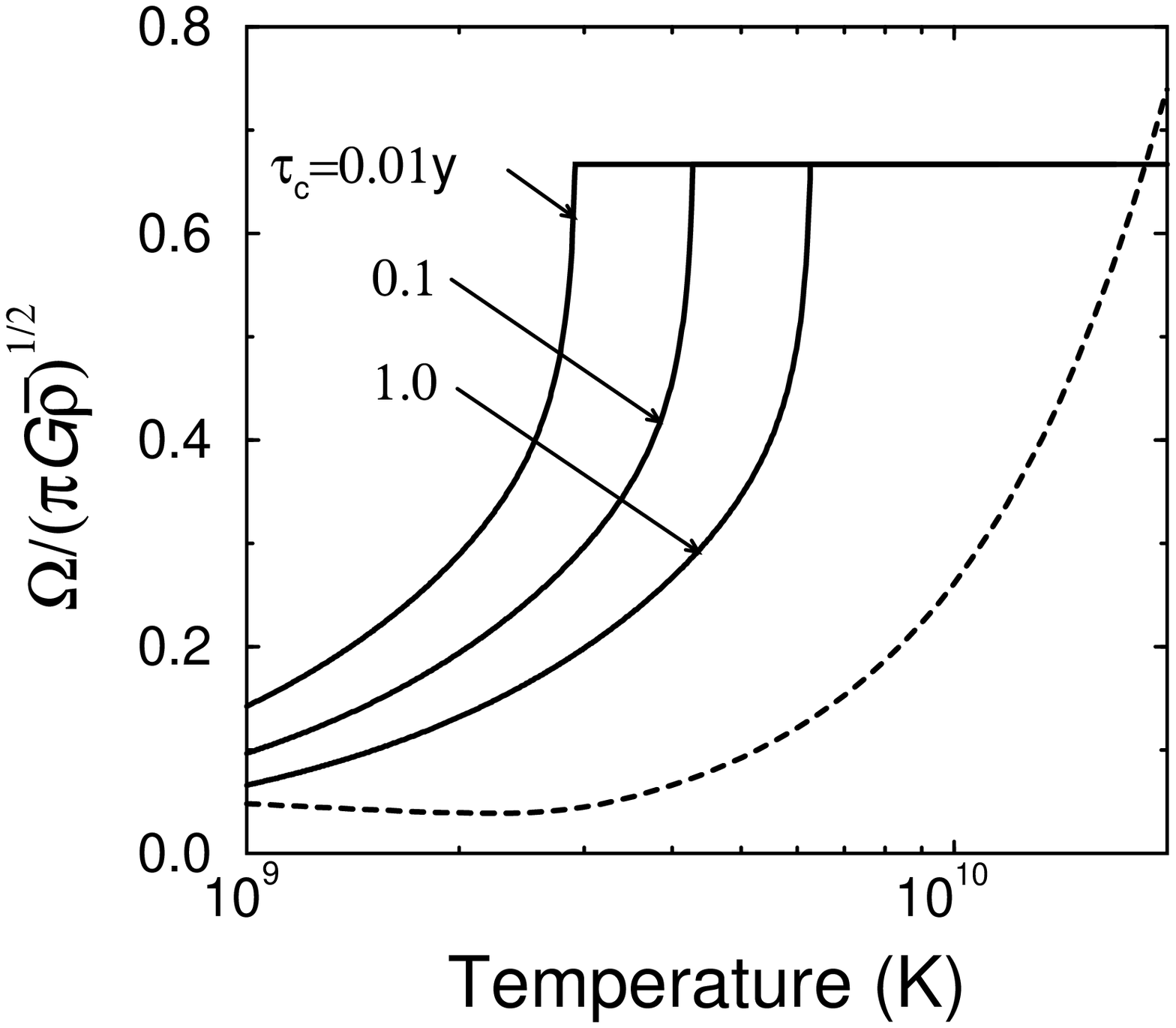,height=2.0in}}
\vskip 0.3cm
\caption{Angular velocity evolution due to gravitational radiation
emission using several values for the timescale $\tau_c$
on which the star cools.
\label{fig4}}
\efig

\section{Gravitational waveforms}
\label{sectionIV}

As the {\it r-}mode grows and evolves it emits gravitational radiation.
In this section we calculate the waveforms for the gravitational wave
strain $h(t)$ and its Fourier transform $\tilde{h}(f)$ that are
produced by this {\it r-}mode instability.  These are the quantities
that can be measured by the gravitational wave detectors now under
construction (LIGO, VIRGO, GEO, etc.).  During the non-linear
saturation phase of the {\it r-}mode evolution, gravitational
radiation controls the dynamics.  In this case $\tilde{h}(f)$ turns
out to be independent of the details of the evolution, and thus can be
determined quite generally.  An evolutionary model {\it is} needed,
however, to determine $df/dt$, the time dependence of the various
quantities, and the initial and final frequencies of the saturation
phase.  In this section we present the general model independent
derivation of $\tilde{h}(f)$ for the non-linear saturated phase of the
evolution.  We also give expressions for the gravitational wave strain
that apply to the early phases of the evolution using the simple model
discussed in Sec.~\ref{sectionIII}.

The frequency-domain gravitational waveform

\beq
\label{4.1}
{\tilde h}(f) \equiv \int_{-\infty}^{\infty}\, e^{2\pi i f t}h(t)\, dt 
\eeq

\noindent is determined completely by the assumption that the angular
momentum radiated as gravitational waves comes directly from the
angular momentum of the star.  This assumption is expected to be
satisfied during the non-linear saturated phase of the evolution, but
not during the early evolution when the mode is growing exponentially.
This derivation is based on Blandford's analysis (as discussed in
\cite{300yrs}) of white dwarf collapse to a neutron star which is
halted by centrifugal forces (see also \cite{schutz89,thorneIAU}).
Such a star can only collapse to a neutron star by shedding its excess
angular momentum through gravitational waves.  In that situation as in
the non-linear saturation phase of the {\it r-}mode evolution,
gravitational radiation determines the rate at which angular momentum
leaves the system, and this in turn determines the rate at which the
frequency of the radiation evolves with time.

In the stationary phase approximation (which is always valid for a
secular instability) the gravitational wave strain $h(t)$ is related
to its Fourier transform $\tilde{h}(f)$ by

\beq 
|h(t)|^2=|\tilde h(f)|^2 \left|{df\over dt}\right|.
\label{4.2} 
\eeq 

\noindent Throughout this discussion we treat $h(t)$ as a complex
quantity with purely positive frequency.  For the $l=2$ mode of
primary importance here, the mode frequency is $\omega = {4\over 3}
\Omega$, or $f = {2\over 3\pi} \Omega $, where $f$ is the frequency of
the emitted gravitational waves measured in Hz.  Assuming the star is
uniformly rotating, its angular momentum is $J = I \Omega$, where $I$
is the star's moment of inertia.  The moment of inertia is fairly
independent of angular velocity (especially at small angular
velocities where most of the detectable signal from these sources is
likely to originate) and is also fairly independent of the amplitude
of the excited $r$-mode (see Eq.~\ref{3.7}).  Thus, $I$ is reasonably
well approximated by its non-rotating value. Thus

\beq {dJ\over df} = {3\pi\over 2}
I.\label{4.3} 
\eeq 

\noindent The rate at which angular momentum is radiated away by
a source is related to the gravitational wave amplitude by the
expression~\cite{thorne}

\beq 
{dJ\over dt} = 4\pi D^2
{m\over{\omega}}{1\over{16 \pi}} \omega^2 \langle h_+^2 +h_\times^2\rangle
\label{4.4}
\eeq 

\noindent where $h_+$ and $h_\times$ are the amplitudes of the two
polarizations of gravitational waves, $D$ is the distance to the
source, and $\langle\ldots\rangle$ denotes an average over the
orientation of the source and its location on the observers sky.
Using $dt/df = dJ/df (dJ/dt)^{-1}$ and combining
Eqs.~(\ref{4.2})--(\ref{4.4}) we obtain

\beq \langle |\tilde h_+(f)|^2 +
|\tilde h_\times(f)|^2\rangle = {3I\over 4D^2 f}.
\label{4.5}
\eeq

The measured value of $|{\tilde h}(f)|^2$ depends on the orientation
of the source and its location on the detector's sky.  Averaged over
these angles, its value is given by

\beq
\langle |\tilde h(f)|^2\rangle = {\scriptstyle {1\over 5}} 
\langle |\tilde h_+(f)|^2 + |\tilde h_\times(f)|^2\rangle .
\label{4.6}
\eeq

\noindent We are actually interested in the average over source
locations in three-dimensional space, not just the two angles on the
sky.  The spatial average weights more strongly those orientations
that yield stronger signals, effectively increasing
$\langle|\tilde{h}(f)|^2\rangle$ by about ${3\over2}$.  Combining
these results then, the average value of $\tilde{h}$ produced by our
fiducial {\it r-}mode source (with $M=1.4M_\odot$, $D=20$Mpc,
$R=12.5$km) is

\beq \tilde{h}(f)= 5.7\times10^{-25}
 \sqrt{1\mbox{ kHz}\over f}\,{\rm
Hz}^{-1}.\label{4.7} 
\eeq 

\noindent Note that this expression does not depend (in the frequency
domain) on the details of the evolutionary model apart from the upper and
lower frequency cutoffs.  This expression, Eq.~(\ref{4.7}), only
depends on the assumption that the frequency of the mode evolves as
angular momentum is radiated by the star according to Eq.~(\ref{4.3}).
We expect this to be satisfied during the non-linear saturated phase
of the {\it r-}mode evolution, but probably not during the early
linear growth phase of the mode.

To obtain the complete waveforms for a particular evolutionary model,
we start with the usual expression for the gravitational field in
terms of its multipoles~\cite{thorne}.  We average this expression
over angles in the manner described above to obtain

\beq 
h(t)= \sqrt{3\over
80\pi}{\omega^2 S_{22}\over D}.\label{4.8} 
\eeq 

\noindent For the simple two-parameter evolution of the {\it r-}mode
instability described in
Sec.~\ref{sectionIII},  this expression can be simplified using
Eq.~(\ref{3.9}) to

\beq h(t)= 
4.4\times 10^{-24} \alpha
\left({\Omega\over \sqrt{\pi G \bar{\rho}}}\right)^3 \left({20\,{\rm
Mpc}\over D}\right).\label{4.9} 
\eeq 

\noindent This simple evolutionary model also gives a simple formula
for the frequency evolution.  During the non-linear saturation phase
of the evolution Eq.~(\ref{3.17}) can be written as

\beq
\label{4.10}
{df\over dt}\approx -1.8\kappa\left({f\over1\mbox{ kHz}}\right)^7\,{\rm Hz/s},
\eeq

\noindent where we have assumed that $\kappa Q\ll 1$.  The time
for the gravitational wave frequency to evolve to its minimum value
$f_{\min}$ is obtained by integrating Eq.~(\ref{4.10}):

\beq
\label{4.11}
t \approx {1.0 \over \kappa} \left({120{\rm Hz}\over f_{\min}}\right)^6 
{\rm y}.
\eeq

Analogous model dependent expressions can also be derived for the
early linear phase of the evolution.  During this period the amplitude
$\alpha$ grows exponentially on the gravitational timescale according
to Eq.~(\ref{3.14}), while the frequency of the mode changes extremely
slowly according to Eq.~(\ref{3.15}).  Solving these equations approximately
gives

\beq
{df\over dt} \approx -{2.7 \alpha^2\over t}
\left({f\over 1\,\, {\rm k Hz}}\right)^3,
\label{4.12}
\eeq

\noindent Using Eq.~(\ref{4.9}) this implies that $\tilde{h}$ during
the linear growth phase is given approximately by

\beq
\label{4.13}
\tilde{h}(f)\approx
4.7\times 10^{-25}\sqrt{t} 
\left({f\over 1\,\,{\rm k Hz}}\right)^{3/2} {\rm Hz}^{-1}.
\eeq
\noindent where $t \equiv t(f)$ is obtained from Eq.~(\ref{4.12}).
The factor $f^{3/2}$ that appears in Eq.~(\ref{4.13}) is essentially
constant, being given by the initial mode frequency as determined by
the initial angular velocity of the star.  Since this factor is
essentially constant during the linear evolution phase, this implies
that $\tilde{h}(f)$ grows as $\sqrt{t}$.  The duration of the initial
linear growth phase can be estimated from the solution for the
amplitude $\alpha$:

\beq
\label{4.14}
\Delta t = 1.5\times 10^3 \left({1\,\,{\rm k Hz}\over f}\right)^6
{\ln{(\sqrt{\kappa}/\alpha_o)}\over \ln{(10^6)}}\,{\rm s},
\eeq

\noindent where $\alpha_o$ is the initial size of the perturbation.
During this interval, the mode frequency decreases by an amount of
order $0.1 \kappa\, {\rm Hz}$. This is the width of the initial spikes
shown in Figs. 7 and 8.  The maximum amplitude achieved by $\tilde{h}$
can also be determined from Eqs.~(\ref{4.13}) and (\ref{4.14}):

\beqa
\max(\tilde{h})\approx 1.8\times 10^{-23}
&&\left({1\,\,{\rm k Hz}\over f}\right)^{9/2}\nonumber \\
&&\times
\left[{\log{(\sqrt{\kappa}/\alpha_o)}\over \log{(10^6)}}\right]^{1/2}
{\rm Hz}^{-1}.
\label{4.15}
\eeqa

\noindent This expression for $\max({\tilde{h}})$ is fairly
insensitive to the duration of the growth phase, and as well to the
exact point at which the transition to the non-linear saturation phase
occurs.  The value of $\max(\tilde{h})$ {\it is} rather sensitive to
our expression for $df/dt$ however.  If $df/dt$ were to differ during
the linear growth phase from the expression used here by any small
effect (such as non-linear modifications of the frequency of the mode)
then the the resulting change on $\max(\tilde{h})$ could be large.
The analytical expressions given here,
Eqs.~(\ref{4.13})--(\ref{4.15}), do however accurately represent (to
within a few percent) the exact numerical solutions to the equations
for our simple model in this region.

Figure~\ref{fig5} illustrates our full numerical solutions for the time
dependence of the gravitational wave amplitude $h(t)$ for several
values of the parameters $\alpha$ with $\kappa=1$.  Figure~\ref{fig6}
illustrates the time dependence of $h(t)$ for various values of
$\kappa$ with $\alpha = 10^{-6}$.  All of these curves represent the
gravitational radiation emitted by a neutron star initially spinning
with angular velocity $\Omega=\Omega_K={\scriptstyle {2\over
3}}\sqrt{\pi G \bar{\rho}}$.  In this section and the next we
terminate the evolution once the star has cooled to $10^9$K.  Below
this temperature the evolution will be significantly affected by
mechanisms not dealt with in this paper, such as the superfluid
transition, the re-heating of the star by viscous dissipation in the
mode, and dissipation mechanisms (e.g. plate tectonics) associated
with the rapidly forming crust.

\bfig
\centerline{\psfig{file=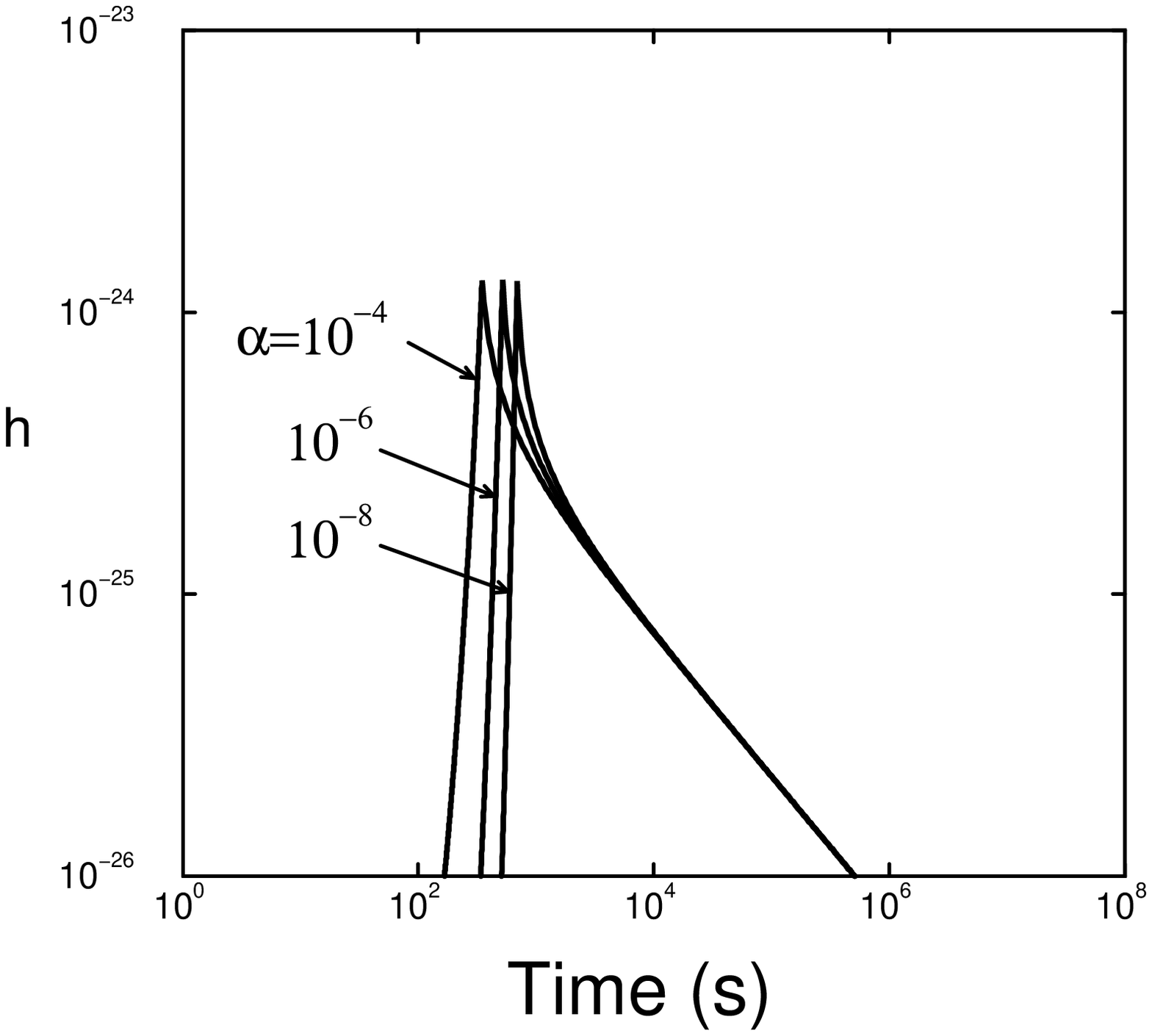,height=2.0in}}
\vskip 0.3cm
\caption{Time dependence of the gravitational wave amplitude
$h(t)$ for a detector located at $D=20\,$Mpc.  The peak
amplitude is very insensitive to the initial size of the
perturbation $\alpha$.\label{fig5}}
\efig

\bfig
\centerline{\psfig{file=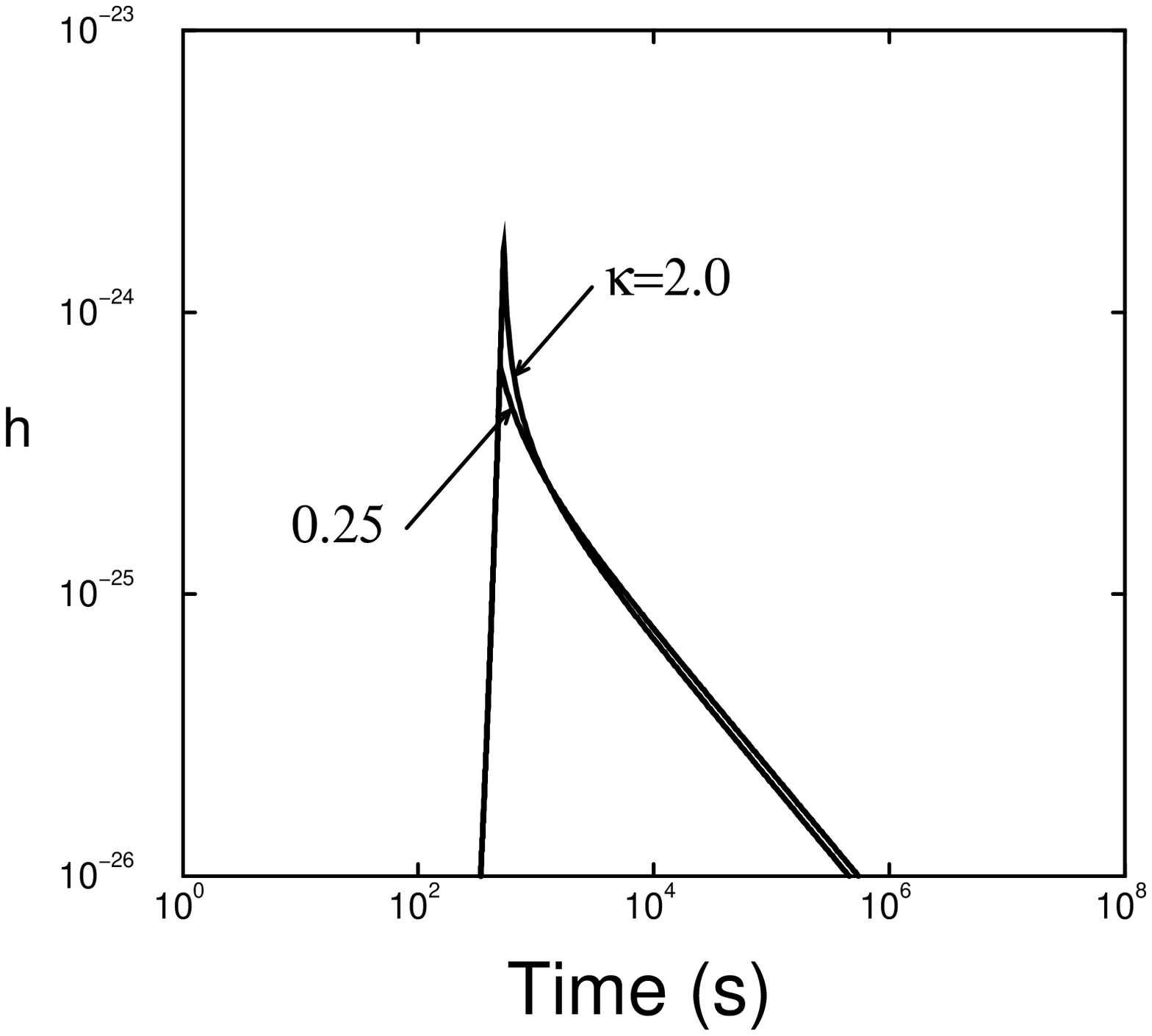,height=2.0in}}
\vskip .3cm
\caption{Time dependence of the gravitational wave
amplitude $h(t)$.  The late time amplitude is rather insensitive
to the non-linearity parameter $\kappa$.\label{fig6}}
\efig

Figs.~\ref{fig7} and~\ref{fig8} illustrate the frequency dependence of
$\tilde{h}(f)$ for a $1.4M_\odot$ neutron star located at $20\,$Mpc
using the same values of $\alpha$ and $\kappa$ used in
Figs.~\ref{fig5} and \ref{fig6}.  Figure~\ref{fig7} illustrates that
$\tilde{h}(f)$ is remarkably insensitive to the initial size of the
perturbation $\alpha$.  The sharp vertical spikes appearing at the
high-frequency ends of these curves are due to the extremely
monochromatic gravitational waves emitted during the linear growth
phase.  The structure of this spike in our model is accurately
described by Eqs.~(\ref{4.13})--(\ref{4.15}), but it is not clear that
this spike is a robust feature of our model.  During the phase of the
evolution that produces the spike, the amplitude of the mode is quite
small except for a period of about one minute.  Thus the total amount
of radiated energy and angular momentum contained in this spike is
quite small. The spike is not likely to play an important role in the
detection of these sources even if it is a real feature of the
$r$-mode instability.

\bfig
\centerline{\psfig{file=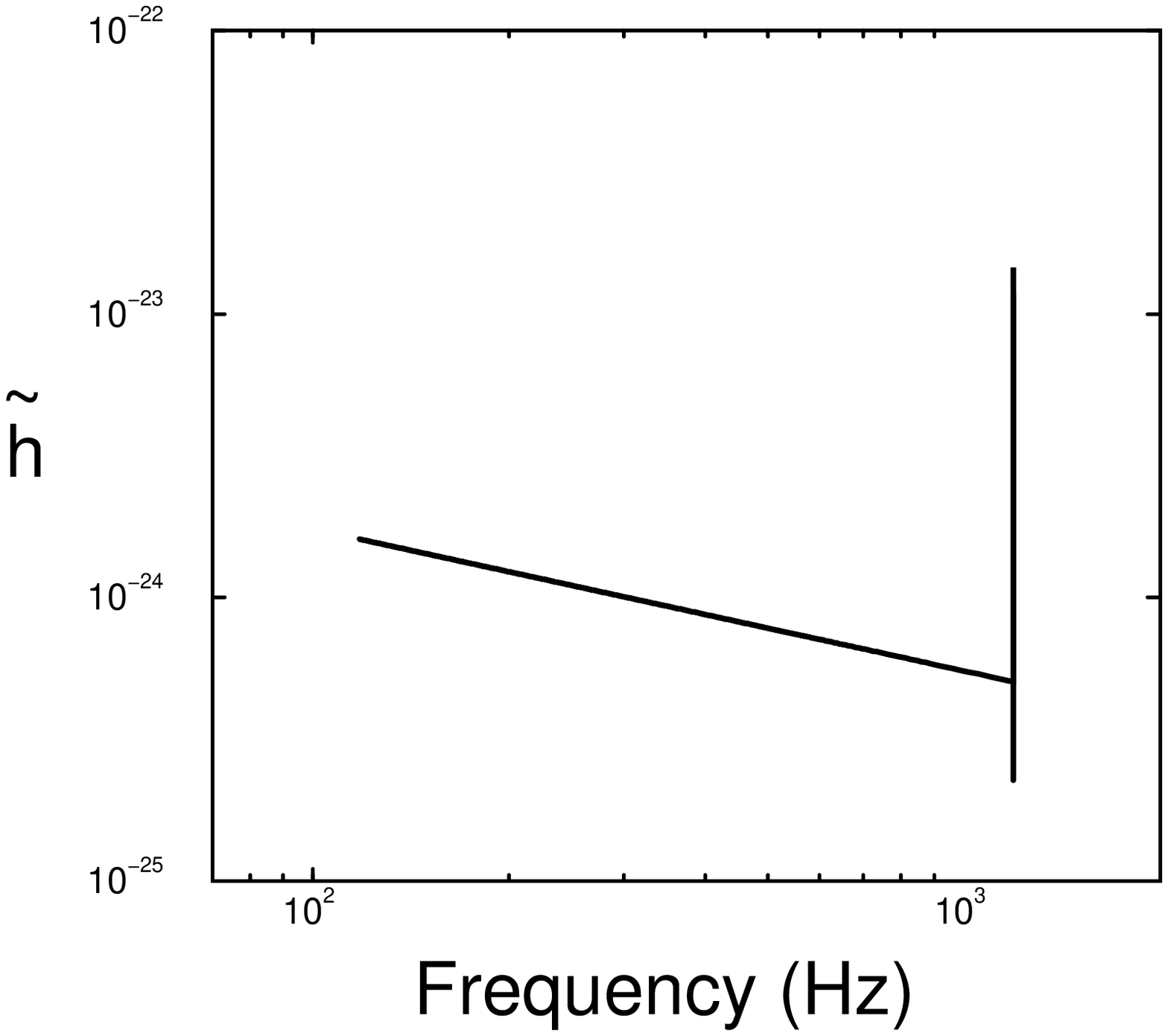,height=2.0in}}
\vskip 0.3cm
\caption{Frequency dependence of the gravitational wave amplitude
$\tilde{h}$ for a source located at $20\,$Mpc.
\label{fig7}} 
\efig 

\bfig
\centerline{\psfig{file=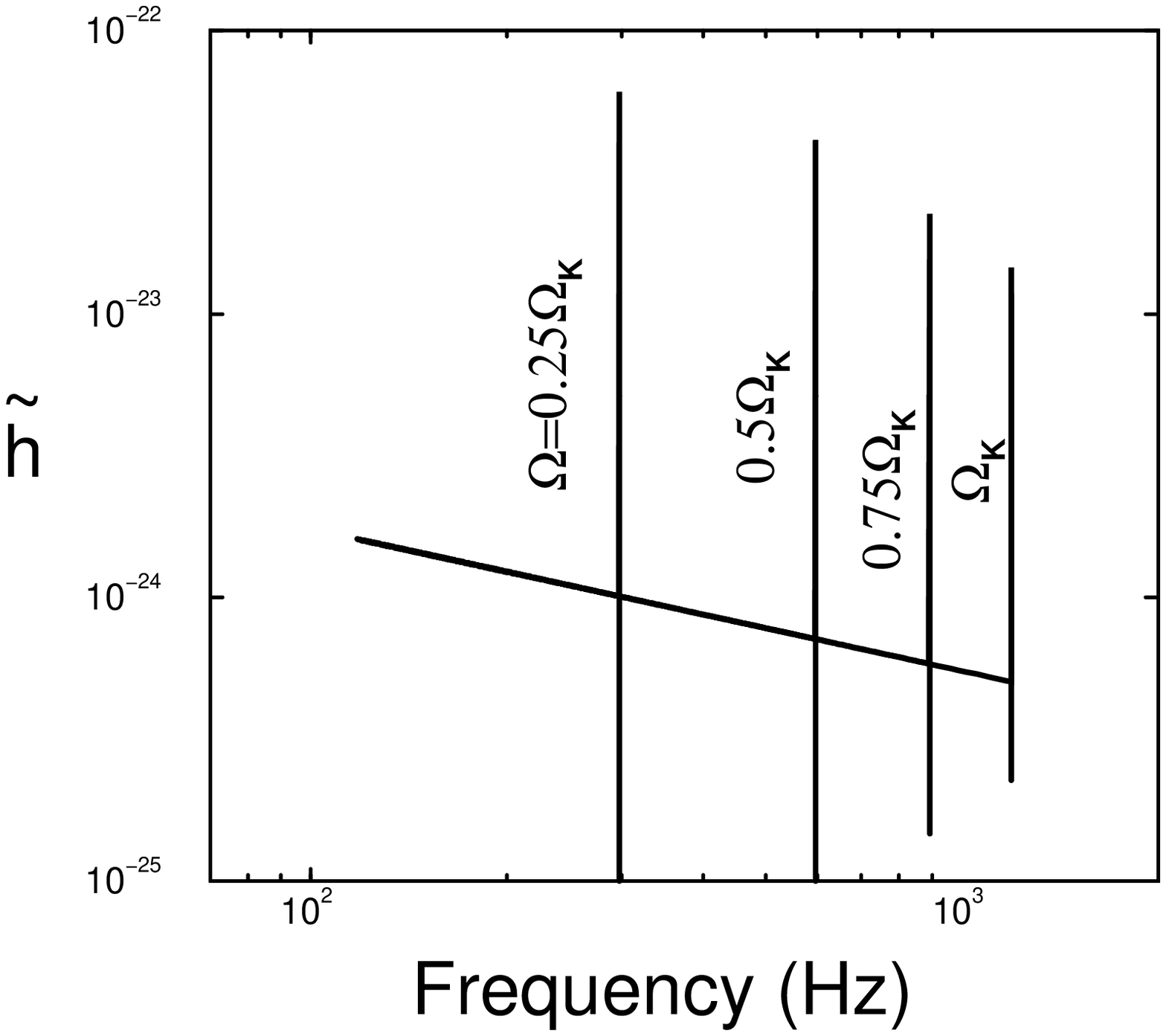,height=2.0in}}
\vskip 0.3cm
\caption{Frequency dependence of the gravitational wave amplitude
$\tilde{h}$ for various values of the initial angular velocity of the
star.\label{fig8}}
\efig

\section{Detectability}
\label{sectionV}

In this section we discuss the detectability of the gravitational
radiation emitted by young neutron stars spinning down due to the
$r$-mode instability.  This radiation might be detected as strong
sources from single spindown events or as a stochastic background made
up of many weaker sources.

\subsection{Single Sources}

First we estimate the signal-to-noise ratio $S/N$ for a single source
located at a distance $D$, chosen to be large enough so that there is
likely to be a reasonable event rate (say a few events per year). This
distance $D$ must be large enough then to include several thousand
galaxies (assuming that the observed supernova rate is comparable to
the neutron star formation rate).  Thus we take this fiducial distance
to have the value $D=20\,$Mpc~\cite{thorne78}, the approximate
distance to the Virgo cluster of galaxies.  We estimate the optimal
value of $S/N$ that could be obtained by matched filtering.
Because matched filtering is probably not feasible for these sources,
this estimate provides only an upper limit of what might be achieved.
We briefly discuss two more realistic search strategies based on
barycentered Fourier transforms of the data~\cite{bccs}.

Using matched filtering, the power signal-to-noise ratio $(S/N)^2$ of
the detection is given by

\beq
\left({S\over N}\right)^2 
= 2\int_0^{\infty} \, { {|{\tilde h}(f)|^2 \over {S_h(f)} } }\,\, df
\label{5.1}
\eeq 

\noindent where $S_h(f)$ is the power spectral density of the detector
strain noise.  The constant in front of the integral in
Eq.~(\ref{4.5}) is 2 (instead of 4 as in Ref.~\cite{cutler-flanagan})
because our $h$ is complex (with purely positive frequency).
Equation~(\ref{5.1}) can also be written

\beq
\label{5.2}
\left({S\over N}\right)^2
=2\int {df\over f}\left({h_c\over h_{\rm rms}}\right)^2.
\eeq

\noindent Here the rms strain noise $h_{\rm rms}$ in the detector
is given by

\beq
\label{5.3}
h_{\rm rms} \equiv \sqrt{f\,S_h(f)},
\eeq

\noindent where $S_h(f)$ is the power spectral density of the noise,
and the characteristic amplitude $h_c$ of the signal is defined
by~\cite{300yrs}

\beq
\label{5.4}
h_c \equiv h\sqrt{f^2 \left|{dt\over df}\right|}.
\eeq

\noindent The quantity multiplying $h$ on the right side of
Eq.~(\ref{5.4}) is generally interpreted as the number of cycles
radiated while the frequency changes by an amount of order $f$.  This
interpretation is correct as long as the frequency evolution is very
smooth.

However, our evolutionary model of the frequency evolution contains a
discontinuity as the mode stops linearly evolving and saturates.
Consequently the actual number of cycles spent near the initial
frequency is far fewer than indicated by Eq.~(\ref{5.4}).  The
quantity $h_c$ is a useful estimator of the effective filtered
amplitude of a signal because the {\it integral} of $(h_c/h_{\rm
rms})^2$ always gives the optimal $(S/N)^2$.  Therefore a spike in
$h_c$ must be interpreted with some caution---the peak value of the
spike is useless unless one also knows the bandwidth of the spike.

In Figure~\ref{fig9} we plot $h_c$ versus frequency, superimposed on
$h_{\rm rms}$ for three LIGO interferometer configurations.  In
the saturation phase (i.e., not including the spike) $h_c$ is well
approximated by

\beq
h_c \approx 5.7\times10^{-22}\left({f\over1\mbox{ kHz}}\right)^{1/2}.
\eeq

\noindent We plot $h_{\rm rms}$ for the LIGO ``first
interferometers''~\cite{science} (which we abbreviate LIGO~I), the
``enhanced interferometers''~\cite{r-d} (LIGO~II), and the ``advanced
interferometers''~\cite{science} (LIGO~III).  The noise power spectral
density for LIGO~I is well approximated by the analytical
fit~\cite{thorne-fit}

\beq
\label{5.5}
S_h(f)={S_o\over 3}\left[\left({f_o\over f}\right)^4+2\left({f\over
f_o}\right)^2\right]
\eeq

\noindent where $S_o=4.4\times 10^{-46}$ Hz${}^{-1}$ and $f_o=175\,$Hz.  For
LIGO~II we construct the approximation

\beq
\label{5.6}
S_h(f)={S_o\over 11}\left\{2\left({f_0\over f}\right)^{9/2}
+{9\over2}\left[1+\left({f\over f_0}\right)^2\right]\right\}
\eeq

\noindent where $S_o=8.0\times 10^{-48}$ Hz${}^{-1}$ and $f_o=112$Hz.  For
LIGO~III the noise spectral density is well approximated
by~\cite{hughes}

\beq
\label{adv}
S_h(f)={S_o\over 5}\left\{\left({f_0\over f}\right)^4
+2\left[1+\left({f\over f_0}\right)^2\right]\right\}
\eeq

\noindent where $S_o=2.3\times 10^{-48}$ Hz${}^{-1}$ and $f_o=76\,$Hz.

\bfig
\centerline{\psfig{file=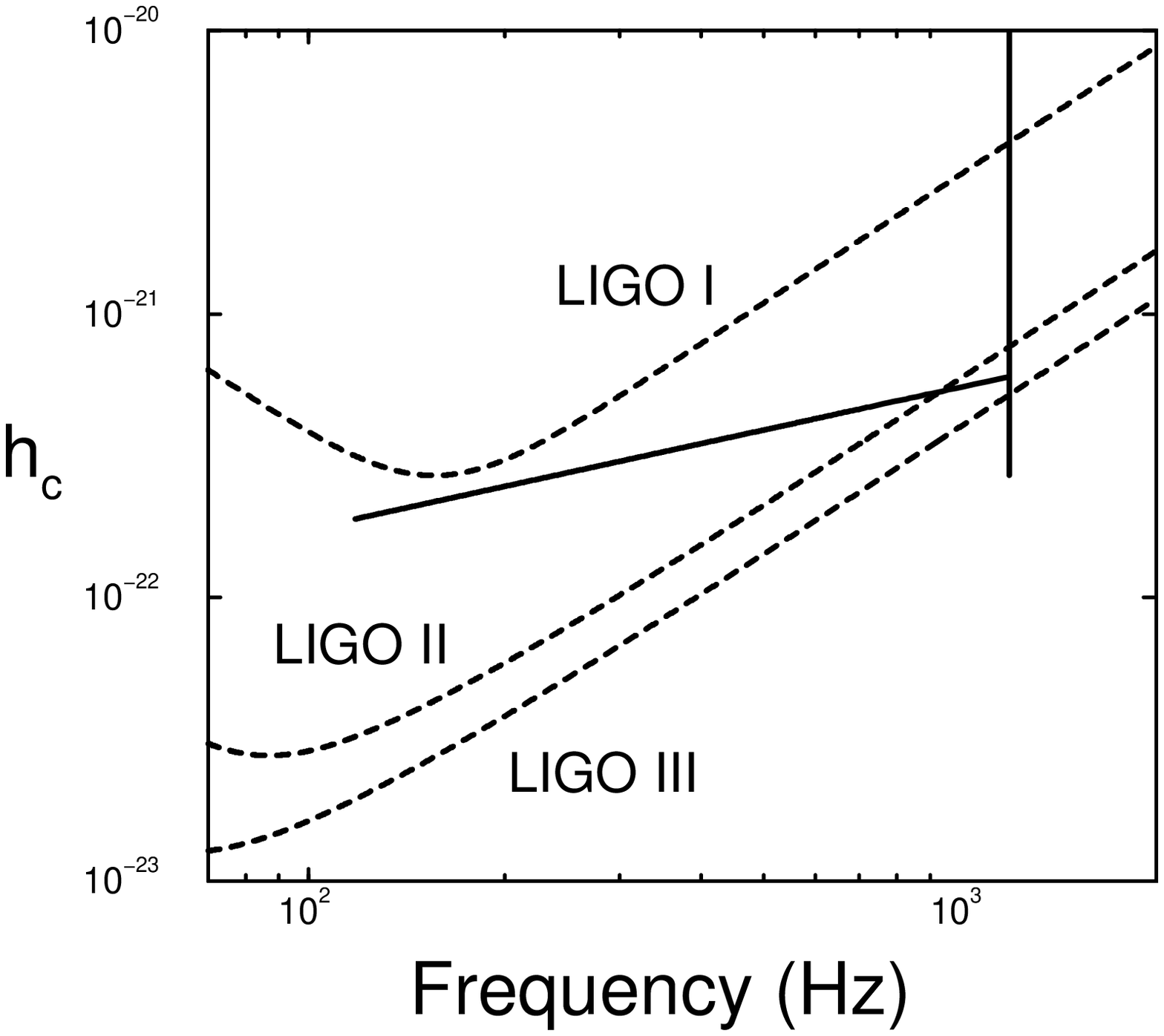,height=2.0in}}
\vskip 0.3cm
\caption{Characteristic gravitational wave amplitude $h_c$ (solid curve)
compared to the noise amplitude $h_{\rm rms}$ (dashed curves) for LIGO.
\label{fig9}} 
\efig

Most of the contribution to $S/N$ in Eq.~(\ref{5.2}) comes from the
saturation phase of the evolution which is largely model independent
as discussed earlier.  Given a detector noise curve, $S/N$ is thus
independent of most details of the waveform except the final frequency
of the neutron star.  We therefore have

\beq
\label{5.7}
\left({S\over N}\right)^2 
= {9I\over 10D^2} \int_{f_{\min}}^{f_{\max}} \,{df\over f\,S_h(f)}.
\eeq

\noindent The minimum frequency $f_{\min}$ reached by the $r$-mode
evolution is about $120$ Hz.  At frequencies slightly larger than
$f_{\min}$, the LIGO~II noise is dominated by photon shot noise. If we
ignore the other noise components, $S_h(f)$ becomes

\beq
S_h(f) = 2.6 \times 10^{-52} f^2.
\label{5.8}
\eeq

\noindent For $f_{\max}\gg f_{\min}$ the integral is dominated by the
lower cutoff.  Thus, the LIGO~II $S/N$ is given approximately by

\beq
{S\over N} \approx 8.8 \left( {I\over{10^{45}{\rm cgs}}}\right)^{1/2} 
\left({D\over{20\,{\rm Mpc}}}\right)^{-1} 
\left({{f_{\min}}\over{120\,{\rm Hz}}}\right)^{-1}.
\label{5.9}
\eeq

\noindent Including the other components of the noise decreases $S/N$
somewhat.  For the numerical evolutions of our simple model with
$\kappa=1.0$ which terminate at $10^9\,$K, we find we find $S/N =
1.2$, $7.6$, and $10.6$ for LIGO~I, II, and III respectively at $D=20$
Mpc.  These last three results scale with $I$ and $D$ just as in
Eq.~(\ref{5.9}), but the dependence on $f_{\min}$ is now more
complicated.  The contribution to $S/N$ from the high frequency spike
in our model is $0.6$ for LIGO~II, and about 0.1 for LIGO~I.  While
the height of the spike in $h_c$ may not be a robust feature of our
simple model, the contribution that this spike makes to the overall
$S/N$ probably is.  These numbers indicate that the gravitational
radiation from the {\it r-}mode instability is a probable source for
LIGO~II if some near-optimal data analysis strategy can be developed.
It appears unlikely that the radiation from the high frequency spike
will be detectable.

Given the recent discovery of the ultrafast young pulsar
PSR~J0537--6910 in the Large Magellanic Cloud~\cite{lmc}, it is
interesting to examine the effect on detectability of a relatively
high superfluid transition temperature.  If the initial period of
PSR~J0537--6910 was about 7~ms (as extrapolated from the braking
indices of typical young pulsars), that could indicate a superfluid
transition temperature of about $2\times10^9\,$K and final
gravitational-wave frequency of about 200~Hz.  Cutting off our
simulations at this point in the evolution, we obtain for LIGO~II
the $S/N$ of about 5.

It is clear from Fig.~\ref{fig9} that the first-generation LIGO and
VIRGO detectors will not see $r$-mode events from the Virgo cluster.
Their sensitivity is a factor of about 8 worse than the enhanced
detectors.  We have also considered the possibility that
GEO~\cite{geo600} might detect these sources by using narrow-banding,
where it can improve its sensitivity in a restricted frequency range
at the expense of worse sensitivity elsewhere.  However, for the kind
of broad-spectrum signal produced by the $r$-mode instability,
narrow-banding is in fact neutral: the gain of signal-to-noise ratio
in the selected band is just compensated by the loss over the rest of
the spectrum.  So GEO is not likely to see these signals either.  Nor
will advanced resonant detectors: at their frequencies and in their
relatively narrow bandwidths, there is just not enough power in these
signals if the sources are in the Virgo cluster.  For example, the
proposed GRAIL detector~\cite{grail} operating at its quantum-limited
sensitivity ($S_h = 1.6\times 10^{-48}\rm\;Hz^{-1}$) between 500 and
700~Hz would have $S/N \approx 1$ for a source at the distance of the
Virgo cluster.

Matched filtering using the year-long waveform templates that would be
needed to describe these sources completely could yield the
signal-to-noise ratios quoted above.  However, this is not a practical
strategy for this type of signal due to the prohibitively large number
of templates that would be needed to parameterize our ignorance of
these sources and the resulting high computational cost of filtering
the data with these templates.  Other strategies equivalent to
combining the results of shorter template searches might well be
computationally feasible, although they would obtain less than the
optimal $S/N$.

The barycentered fast Fourier transform (FFT) technique that has been
designed to search for nearly periodic signals~\cite{bccs} might well
provide one such method.  Figure~\ref{fig10} shows the spindown age

\beq
\tau_{\rm sd}= -f\,{dt\over df}
\eeq

\noindent for a neutron star spinning down due to the $r$-mode
instability.  This quantity provides a reasonably good estimate of the
time spent by the evolving star in the saturated non-linear phase, but
it is not a good estimate of the amount of time spent during the
linear growth phase (for the reasons outlined above).  During the
saturated phase $\tau_{\rm sd}$ is given approximately by

\beq
\tau_{\rm sd} \approx {580\over \kappa}\left({1\mbox{ kHz}\over f}
\right)^6\mbox{ s}\,
\approx {6t\over \kappa}.
\eeq

\noindent 
The signal becomes quite monochromatic after about one day, in the
sense that the spindown timescale is long compared to the inevitable
daily modulation of the signal due to the motion of the Earth.  Thus
the search techniques for periodic sources should work well.  In many
cases the supernova will be observed, yielding the location of the
source and allowing a search over spindown parameters only, a
{\it directed spindown search.}

The most straightforward way to conduct a directed spindown search is
to search over generic spindown parameters as discussed by Brady and
Creighton~\cite{brady-creighton}.  This involves re-sampling the data
so as to render sinusoidal a signal with arbitrary (smooth) frequency
evolution, then Fourier transforming it.  The signal frequency
evolution (which determines the re-sampling) is modeled as

\beq
\label{5.13}
f(t)=f_0\sum_{n=0}^N {1\over n!}\left({t\over \tau_{n}}\right)^n,
\eeq

\noindent where $f_0$ is the frequency at the beginning of the FFT,
$t$ is the time measured from the beginning of the FFT, and $\tau_n$
are expansion parameters with $\tau_1=\tau_{\rm sd}$.  The number $N$
of spindown parameters needed is set by the requirement that the
frequency drift due to the next term in the series~(\ref{5.13}) be
less than one frequency bin of the FFT (which is in turn determined by
the integration time $\tau_{\rm int}$).  This implies (assuming
$\tau_n\approx\tau_{\rm sd}$) that

\beq
\tau_{\rm int}^{N+1}
\approx {N!\over f_0}\tau_{\rm sd}^N.\label{5.14}
\eeq

One must choose a set of points in the spindown parameter space for
which to perform the re-sampling and FFT.  Too few points and one
misses signals by re-sampling at the wrong rate; too many and the
computational cost of performing all the FFTs becomes prohibitive if
data analysis is to be performed ``on-line,'' i.e.\ keep up with data
acquisition.  The points in parameter space are chosen using a metric
which relates distance in parameter space to loss of
$S/N$~\cite{bccs}.  The integration time, which is set so as to
optimize the sensitivity of an on-line search for the (fixed)
computational power available, is far shorter than a year even for a
teraflop computer.  Therefore such a search would achieve at best a
fraction of the optimal $S/N$~\cite{brady-creighton}.  It is possible
that a hierarchical version of this strategy could be developed, in
which the best candidates from a year of shorter FFTs are somehow
combined to give an improved confidence level.  Developing such a
strategy would require extensive further investigation.

One way to increase the sensitivity of a directed spindown search
would be to constrain the spindown parameter space by taking advantage
of whatever information we have about the source from physical models.
In practice we do not have and probably will never have completely
reliable models for this type of system.  The phenomenological model
presented here has many physical assumptions that may prove to be
inadequate.  For instance, in the saturation phase it is highly
unlikely that the star can be represented simply as a uniformly
rotating equilibrium configuration plus a linear perturbation.  When
the mode reaches saturation, the mean velocity perturbation is
comparable to the rotation rate of the star.  Thus in the saturation
phase, the velocity field may develop complicated nonlinear structures
(such as the cyclones on Jupiter) that produce gravitational radiation
involving many different multipoles.  Long-lived non-axisymmetric
structures requiring higher-order modes are seen in the density
perturbations of simulations of star formation when there is enough
angular momentum~\cite{tohline}.  Although the physics is very
different in such simulations, a priori we see no reason to believe
that the velocity analogues of these structures are not formed by the
$r$-mode instability.

However, not all of the physics affects the waveforms.  Some kind of
phenomenological model of the signal (as opposed to the star) could be
enough to substantially reduce the volume of spindown parameter space
to be searched.  For a reasonable model of the neutron star, all of
the terms in the expansion for $df/dt$ might be determined from a
relatively small number of phenomenological parameters.  Although
quite crude, the model of {\it r-}mode instability gravitational
waveforms provided here has several features which should be fairly
robust: the form of $\tilde{h}(f)$ during the spindown phase, the
approximate frequency range of the expected radiation, the approximate
timescale for the spindown to occur, etc.  Presumably these robust
features can be used to reduce considerably the volume of general
spindown parameter space which need be searched.

\bfig
\centerline{\psfig{file=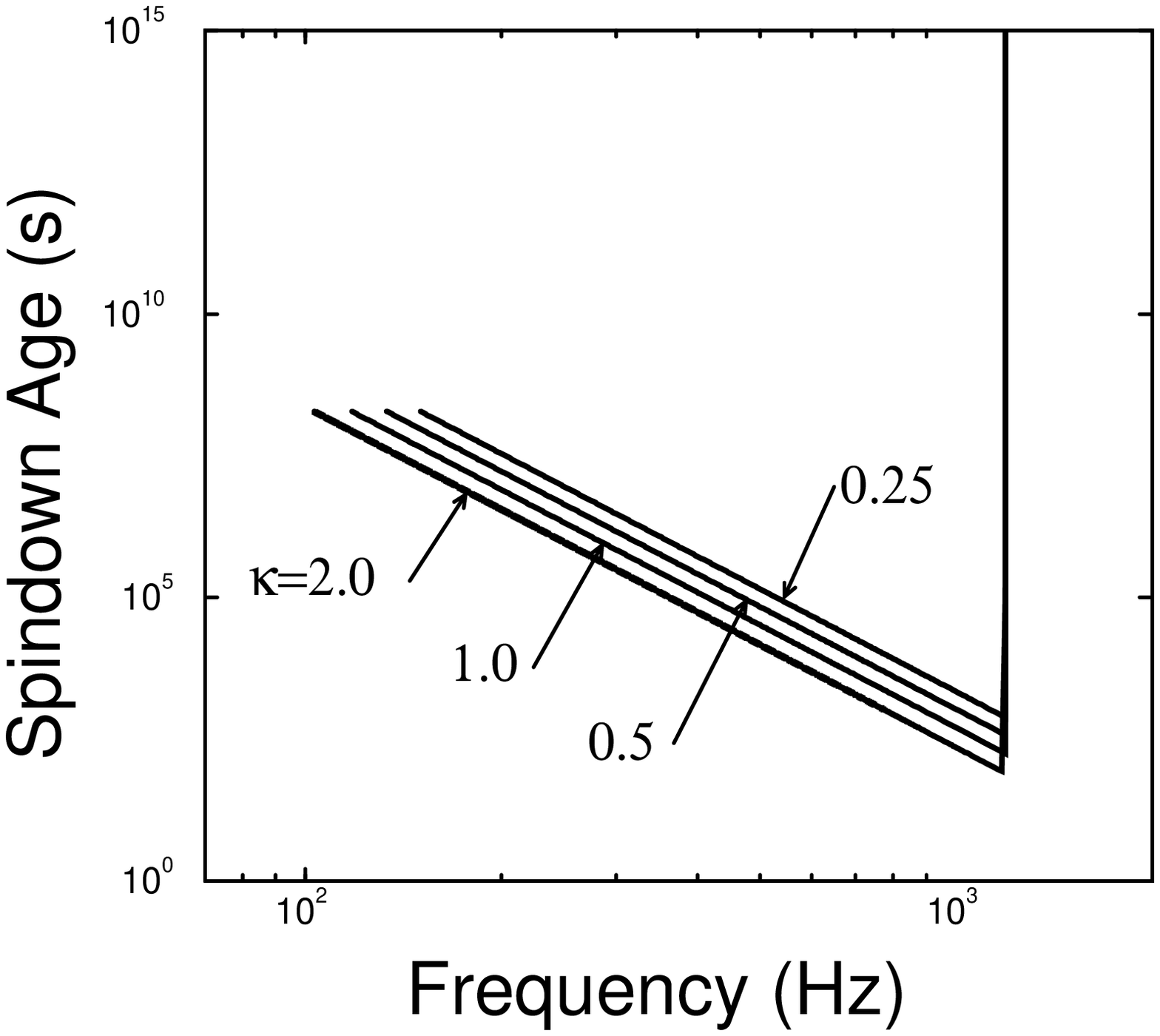,height=2.0in}}
\vskip 0.3cm
\caption{Spindown age for an {\it r-}mode-driven young neutron star for
different values of the nonlinearity parameter $\kappa$.
\label{fig10}}
\efig


\subsection{Stochastic Background}

We now consider the gravitational-wave stochastic background generated
by spin-down radiation from neutron star formation throughout the
universe.  A stochastic background is detected by looking for
correlations in the response of two or more detectors.  The
sensitivity of the network to the background drops rapidly for
gravitational wave frequencies much higher than the inverse
light-travel time between detectors. For the present application, the
important networks are therefore the VIRGO-GEO pair (high frequency
cut-off $f_{\rm cut}\approx 400$ Hz) and the Washington-Louisiana pair
of LIGO detectors ($f_{\rm cut} \approx 100$ Hz).  The gravitational
radiation generated by the {\it r-}mode spin-down process has
significant power at these relatively low-frequencies, $f
\approx 100-400$ Hz. In addition the cosmological redshift (with $z \approx
1-4$) of these sources has the beneficial effect of shifting much of
the radiation into the detectable band.

Neutron stars have presumably been formed since the beginning of star
formation.  If each neutron star formation event converts a reasonable
fraction of a solar mass into gravitational radiation via the $r$-mode
instability, then the sum of this radiation constitutes a random
background that may be detectable by LIGO~III.  We now make a rough
estimate of the spectrum and detectability of this background
radiation.  A more detailed analysis is being carried out by Vecchio
and Cutler~\cite{vecchio-cutler}.

The spectrum of the gravitational wave background is typically
represented by the following dimensionless quantity:

\beq
{\bf \Omega}_{\rm gw}(f) 
\equiv {1\over{\rho_c}} {{d\,\rho_{\rm gw}}\over{d\,\log f}}, 
\label{e:om_f}
\eeq

\noindent where $\rho_{\rm gw}$ is the energy density in gravitational
waves, and $\rho_c=3c^2H_0^2/3\pi G \approx 1.6\times 10^{-8}
h^2_{100}$ erg/cm${}^3$ is the critical energy density just needed to
close the universe. ($H_0$ is the Hubble constant and $h_{100}$ is
$H_0$ divided by 100 km/s/Mpc.)  The signal-to-noise with which this
background can be detected in a correlation experiment between two
detectors (here assumed to have uncorrelated noise) is given
by~\cite{flanagan,allen}

\beq
\label{snsb}
\left({S\over N}\right)^2 
= {9H_0^4 T\over 50\pi^4} \int_0^\infty \,df 
{\gamma^2(f)\Omega^2_{\rm gw}(f) \over f^6 \,{^1S}_h(f) \,{^2S}_h(f)}.
\eeq 

\noindent where $T$ is the integration
time, $^1S_h(f)$ and $^2S_h(f)$ are the noise spectral densities of
the two detectors, and $\gamma(f)$ is the (dimensionless) overlap
reduction function, which accounts for the fact that the detectors
will typically have different locations and orientations.

We can get a rough estimate of ${\bf \Omega}_{\rm
gw}(f)$ due to the stochastic background of gravitational radiation
from the {\it r-}mode instability as follows.  A $1.4 M_\odot$ neutron
star rotating with Keplerian angular velocity $\Omega_K \approx
{\scriptstyle {2\over 3}} \sqrt{\pi G\bar{\rho}}$ has rotational
kinetic energy $E_{\rm K}\approx 0.025 M_\odot c^2$, two-thirds of
which is radiated as gravitational waves. For now, assume that all
neutron stars are born with angular velocity $\Omega_K$.  For a single
supernova occurring at $z=0$, the spin-down radiation has spectrum $d
E_{\rm gw}/df \approx {4\over 3}E_{\rm K}\, f/f_{\max}^2$, where
$f_{\max} \approx 2\Omega_K/ 3\pi \approx 1400$ Hz.

We are chiefly interested in ${\bf \Omega}_{\rm gw}(f)$ for $f <
f_{\min} \approx 120$ Hz, since that is the range where the pair of
LIGO~II detectors will have their best sensitivity.  Let
$n(z)\,dV\,dz$ be the number of supernovae occurring within co-moving
volume $dV$ and redshift interval $dz$. In the frequency range of
interest then the spectrum of gravitational radiation in the universe
today is given by

\beq {d\,\rho_{\rm gw}(f)\over df} \Delta f =
{ 4E_{\rm K}\over 3f_{\max}^2} 
\int_{z_{\min}(f)}^{z_{\max}(f)}{n(z) {f'\Delta
f'dz \over 1+z}},
\label{e:drhodf}
\eeq

\noindent where $f'\Delta f'= (1+z)^2 f \Delta f$ are the values of
the frequencies as emitted by the source, $z_{\min}(f) \equiv {\rm
max}\{0,f_{\min}/f-1\}$, $z_{\max}(f)\equiv
\min\{z_*,f_{\max}/f-1\}$, and $z_{*}$ corresponds to the
maximum redshift where there was significant star formation.  The
factor $1+z$ in the denominator in Eq.~(\ref{e:drhodf}) accounts for
the redshift in energy of the gravitational radiation.

To evaluate the integral in Eq.~(\ref{e:drhodf}), we must make some
assumption about the rate of Types Ib, Ic, and II supernovae (which
are the ones that leave behind neutron stars). The combined rate in
our galaxy at present is roughly one per 100 years.  If this rate had
been constant, the Galaxy would today contain about $10^8$ neutron
stars.  However at earlier times, for $z$ between 1 and 4, the rate
(with respect to proper time) was significantly higher.  A reasonable
estimate is that our Galaxy contains $3\times 10^8$ neutron stars
today.  Let $n(z)\Delta z$ be the density of neutron star births (per
unit comoving volume) between redshifts $z$ and $z + \Delta z$.  As a
rough, first-cut at this problem, we model $n(z)$ as constant
$n(z)\equiv n_o$ for $0 < z < z_{*}$, where $z_{*} \approx 4$, and
take $n(z) = 0$ for $z > z_{*}$.  In this case the integral in
Eq.~(\ref{e:drhodf}) can be performed to obtain:

\FL
\beq
\label{rspec}
{{\bf\Omega}_{gw}(f)\over Af^2} =
\left\{ \begin{array}{ll}
0,
&\mbox{$0 < f < f_1$},\nonumber \\
(z_* + 1)^2 - \left({f_{\min}/ f}\right)^2 ,
&\mbox{$f_1 < f < f_{\min}$},\nonumber \\
z_*(z_* + 2),
&\mbox{$f_{\min} < f < f_2$},\nonumber \\
(f_{\max}/f)^2 - 1,
&\mbox{$f_2 < f < f_{\max}$},\nonumber \\
0,
&\mbox{$f > f_{\max}$},\\
\end{array} 
\right. 
\eeq

\noindent where $A \equiv { 2n_o E_{\rm K}/(3 \rho_c f_{\max}^2)}$,
$f_1 \equiv f_{\min}/(1+z_*)$  
and $f_2 \equiv f_{\max}/(1+z_*)$.  

We can estimate the value of the constant $n_o$ as follows.  We assume
that the number of neutron stars in a given location is roughly
proportional to the luminosity of the visible matter at that location.
The total luminosity of our Galaxy is $1.4 \times 10^{10} L_\odot$
\cite{binney}, while the number of neutron stars in the Galaxy is
about $3\times 10^8$.  Thus the neutron star mass to total luminosity
of matter ratio is about $0.03 M_\odot/L_\odot$.  The mean luminosity
of the universe is $1.0\times 10^8 h_{100} L_\odot/{\rm
Mpc}^3$~\cite{peebles}.  Thus, the mean mass density of neutron stars
is about $\rho_{\rm ns}\approx3 \times 10^{6} h_{100} M_\odot/{\rm
Mpc}^3\approx 1.1\times 10^{-5} h_{100}^{-1}\,\rho_{\rm c}$.  The
current density of neutron stars is related to the constant $n_o$ by
$\rho_{\rm ns}\approx 1.4 M_\odot z_*n_o \approx 5.6 M_\odot
n_o$. Thus ${2\over 3} n_o E_{\rm K}/\rho_{\rm c} \approx 3.3\times
10^{-8}h_{100}^{-1}$.  For LIGO, the most important range in
Eq.~(\ref{rspec}) is $f_1 < f < f_{\min}$; we can re-write the result for this
portion in the more useful form

\beq
{\bf \Omega}_{\rm gw}(f) \approx 2.4\times 10^{-10} h_{100}^{-1} 
\left[\left({f\over 24\,{\rm Hz}}\right)^2 - 1\right]\,. 
\label{e:omgwf2}
\eeq

\noindent
Evaluating at $f=50$ Hz, with $h_{100}=0.7$ we find ${\bf \Omega}_{\rm
gw}(f=50\,{\rm Hz}) \approx 1.1 \times 10^{-9}$.

Using the above spectrum Eq.~(\ref{rspec}), we evaluate the $S/N$
using Eq.~(\ref{snsb}). For an integration time of $T=10^7$s we find
$S/N = 0.0022$, $0.34$, and $2.6$ for LIGO~I, II, and III
respectively. Since there has been some discussion of building a
second kilometer-size interferometer in Europe, we also consider the
sensitivity to the $r$-mode background of this detector paired with
VIRGO.  We assume that the detectors will be located less than about $
300$ km apart and have the same orientation.  (To model this, we
simply set $\gamma(f) = 1$ in the integral Eq.~\ref{snsb}).  We find
$S/N = 5.6$ assuming both these detectors have LIGO II sensitivity,
$S/N = 0.9$ assuming one detector has LIGO I sensitivity while the
other has LIGO III sensitivity, and $S/N = 23$ assuming both have LIGO
III sensitivity.  Thus we see that detection of the $r$-mode
background will have to wait either for development of ``advanced''
interferometers or for the construction of two nearby detectors with
``enhanced'' sensitivity.  Two nearby ``advanced'' interferometers
could see quite a strong signal.  All the above results on correlation
measurements of the stochastic background assume that magnets will be
eliminated from the LIGO design.  With the current design long-range
correlated magnetic fields from Schuman resonances and lightning
strikes will mimic a stochastic background with ${\bf \Omega}_{\rm
gw}$ approximately $10^{-7}$ to $10^{-9}$~\cite{allen-romano}.

These calculations assumed that all neutron stars are born with spins
near their maximal value $\Omega_K$. It should be clear, however, that
these results for the $S/N$ achievable by the LIGO pair depend only on
the stars being born with spins greater than about $300$~Hz.  Of
course, it could well be that some fraction $F$ of neutron stars are
born with rapid spins, while $(1-F)$ are born slowly spinning. The
values of $S/N$ in this case could be estimated from those given above by
multiplying by $F$.  (See Spruit \& Phinney~\cite{phinney} for a
recent argument that most neutron stars should be born with very slow
rotation rates.)

It has previously been suggested that there could be a detectable
gravitational wave background produced by supernova events
\cite{blair,ferrari}.  The stochastic background due gravitational
radiation from the $r$-modes differs from that previously envisioned
in two important respects.  Previously it was assumed that the
radiation would be emitted in short bursts, forming a random but not
continuous background.  For the $r$-mode background, the long duration
of the emission guarantees that it will be a continuous hum rather
than an occasional pop.  Also, the spectrum from spindown radiation
extends to lower frequencies than had previously been expected from
supernova events.

\section{Discussion}
\label{secVI}

The discovery of a strong source of gravitational waves that is ubiquitous 
and is associated with such interesting objects as supernovae and neutron 
stars inevitably opens up a rich prospect for obtaining astronomical 
information from gravitational wave observations.  We shall discuss here 
some of the more obvious prospects.

{\em Background radiation from $r$-modes.} Pleasantly, the background
requires no detailed modeling of the signal in order to detect
it. However, detection of background radiation from $r$-modes will
probably have to wait for the development of ``advanced''
interferometers (or the construction of two nearby ``enhanced''
interferometers), even if we assume that a large fraction of neutron
stars are born rapidly rotating.  For nearby detectors with LIGO III
sensitivity, $S/N$ is high enough that one could experimentally
measure the spectrum ${\bf \Omega}_{gw}(f)$ with reasonable
accuracy. This might provide very interesting cosmological
information.  For instance, imagine that most neutron stars are born
rapidly rotating.  (This could be verified, at least for neutron stars
born today, by direct LIGO detections of nearby supernovae.) The
background spectrum between $25$ and $50$Hz would then give us direct
information about the star formation rate in the early universe.  And
its spectrum between 50 and perhaps 300~Hz would tell us about the
distribution of initial rotation speeds of neutron stars.

{\em Individual $r$-mode events associated with known supernovae.}
Observations of individual spindown events will be more difficult to
achieve but can be very rewarding.  The easiest case is if the
supernova that leads to the neutron star is seen optically.  This
gives some hint of when the $r$-mode radiation should be looked for,
but more importantly it gives a position.  That reduces the difficulty
of extracting the signal from the detector data stream.  Detection of
the radiation will return the amplitude of the signal, its
polarization, the final spin of the star, and the values of the
parameters of the waveform.  Assuming that the three large detectors
(the two LIGO installations and VIRGO) all detect the signal with
$S/N\approx 8$, the effective combined $S/N$ will be $8\sqrt{3}\approx
14$.  Values of the various parameters will then typically be
determined to 10-20\% accuracy.  The polarization measures the
orientation of the spin axis of the neutron star, which will be
difficult to relate to any other observable.  So it may not be of much
interest.  But the power spectrum of the radiation measures, as we
have seen, the loss of rotational energy of the star, so it is
proportional to $I/D^2$.  If the host galaxy's distance can be
determined to better than the accuracy of the gravitational wave
measurement, which seems likely, then this will provide a direct
measure of the moment of inertia of a neutron star.

There will likely be several detections per year, which will shed
light on a number of uncertainties.  Even if the parameters are only
the Taylor expansion coefficients for the frequency, they will
constrain models of the $r$-mode spindown.  We can expect to get some
information about cooling rates, viscosity, crust formation, the
equation of state of neutron matter, and the onset of superfluidity
(or some combination of these).  We also expect variability from event
to event, due to different initial conditions after gravitational
collapse, such as differential rotation or even the mass of the
neutron star.  Significant differential rotation might affect the
final spin rate of neutron stars; hence any variability in the final
spin speed might shed light on these initial conditions.

If we find we can detect this radiation with confidence, then the
absence of it after a supernova could be a hint that a black hole has
formed instead of a neutron star.  More rarely it might indicate that
the neutron star remained bound in a binary system, and the orbital
modulation of the signal made it impossible to find.

Some supernovae will be especially nearby, and some neutron stars will
be oriented more favorably, so there might be a handful of events with
single-detector $S/N\approx 15-20$.  These will provide particularly good
constraints on the moment of inertia, superfluidity, viscosity, etc.
For these events it may also be possible to trace the waveform back to
its initial stages, and thereby to measure the initial spin of the
collapsed star.  In some cases, a star might be formed with more mass
that it can support after spinning down, and so the $r$-mode spindown
would lead to and be interrupted by collapse to a black hole. This
would probably happen relatively soon after the star is formed; but
for strong events, it might be possible to detect this break.

{\em Using $r$-mode events as supernova detectors.}  Perhaps the most
exciting use of $r$-mode observations would be to identify hidden or
unnoticed supernovae.  If it proves possible to create search
strategies that are efficient enough to detect $r$-mode spindown even
without prior positional information from an optical observation, then
the gravitational wave detectors will become supernova monitors for
the Virgo cluster.  Perhaps as many as half of the supernovae in Virgo
go unnoticed, hidden in thick dust clouds.  LIGO and VIRGO would not
give optical observers advance notice of the supernova: they will
identify a neutron star only a year or so after it was formed.  But
they may be able to locate the position of the event with an accuracy
of better than one arc-second.

This great precision is achieved from the modulation of the signal
produced by the motion of the earth \cite{bfs}.  The angular accuracy
is similar to that achieved for pulsar observations in the radio.
Fundamentally it is the diffraction limit of a gravitational wave
telescope with the diameter of the earth's orbit, because the detector
acts like a synthesis array as it builds up signal along its orbit
around the Sun.  The ratio of the gravitational wavelength of about
1000~km to the diameter of 2~AU is about $3\times10^{-6}$ radians.
This angular accuracy improves with $S/N$ as well, but it can be
degraded by uncertainties in the spindown parameters.  This assumes,
as we have in this paper, that neutron-star cooling takes a year or
more.  If the alternative cooling scenarios are correct and the star
cools in a few days, then the angular accuracy of observations will be
very poor.

For an event at $20$~Mpc, arc-second angular resolution corresponds to
distance resolution of about $60$~pc.  Observations would therefore
not only tell us in which galaxy the event occurred, but even in which
molecular cloud.  Detailed follow-up searches will then be possible
for the expanding nebula, starting perhaps one year after the event.

{\em Other implications for gravitational radiation from supernovae.}
The $r$-mode instability also has implications for our expectations of
other kinds of gravitational radiation from supernova events.
Although detecting supernovae was the goal of initial bar detector
development, it has not been possible before now to provide reliable
predictions of radiation from gravitational collapse.  The $r$-mode
instability is a reliable prediction, but only of radiation long after
the collapse event.  It is clear that a collapse that produces a
rapidly rotating star will be more likely to radiate strongly,
especially if it can reach the dynamical bar-mode instability that is
seen in the lower-density star-formation simulations.  But although
rapid rotation is in some sense natural in gravitational collapse, the
fact that young pulsars like the Crab are slow rotators seemed to
indicate that neutron stars do not form with fast spins.  Now the
$r$-mode instability has provided an explanation for the slow spins of
young pulsars; there is no longer any observational restriction on the
initial spins of neutron stars.

It therefore seems to us much more likely than before that the
gravitational collapse event can also be a strong source of
gravitational waves.  The following scenario seems plausible in at
least the extreme cases where rotation completely dominates the last
stages of collapse.  The collapsed object has so much spin that it
forms a bar shape on a dynamical timescale.  This radiates away
angular momentum in gravitational waves until the star is finally able
to adopt a stable axisymmetric shape.  The strong gravitational
radiation ceases, to be replaced by the developing $r$-mode radiation.
The first burst would be detectable by LIGO~II at the distance of the
Virgo cluster, and a network of such detectors could give a rough
position, which would allow notification of optical astronomers of the
event and multi-wavelength follow-up observations.  In the
gravitational wave data stream, intensive searching for the $r$-mode
radiation would follow.

The unexpected strength of the $r$-mode gravitational wave instability
in young neutron stars has therefore completely changed the prospects
for detection of gravitational waves from supernova events.  Not only
can we look forward to regular detections of the spindown radiation
when enhanced detectors begin operating, but we also now have more
reason to expect strong bursts from the collapse events themselves.
Particularly exciting is the prospect for using networks of LIGO-type
detectors as supernova monitors, able to pinpoint the positions of
hidden supernovae with arc-second precision.  In order to achieve these
goals, however, much work needs yet to be done to eliminate the
uncertainties in our models of young neutron stars and to develop
effective ways of searching for signals of this kind.

\acknowledgments

We thank Patrick Brady, Jolien Creighton, Teviet Creighton, Scott
Hughes, Sterl Phinney, Joseph Romano, and Kip Thorne for helpful
discussions.  This research was supported by NSF grants AST-9417371
and PHY-9796079, by the NSF graduate program, and by NASA grant
NAG5-4093.


\end{document}